\RequirePackage{lineno}
\documentclass[aps,prd,twocolumn,superscriptaddress,showpacs,showkeys,amssymb,floatfix,groupedaddress]{revtex4-1}
\usepackage{booktabs,graphicx,mathrsfs,verbatim,amsmath,units}
\usepackage[caption=false]{subfig}
\usepackage[colorlinks=true,citecolor=blue,filecolor=blue,linkcolor=blue,urlcolor=blue,pdftex]{hyperref}
\usepackage[usenames,dvipsnames]{color}
\usepackage{ulem}
\usepackage{color}
\usepackage{multirow}

\newcommand\figref[1]{Fig.\,\ref{#1}}

\begin{document}	
\title{Measurement of light and charge yield of low-energy electronic recoils in liquid xenon}

\newcommand{\columbia}{\affiliation{Physics Department, Columbia University, New York, NY, USA}}

\author{L.~W.~Goetzke}\altaffiliation{Corresponding author. lukeg@phys.columbia.edu}
\author{E.~Aprile}
\author{M.~Anthony}
\author{G.~Plante}
\author{M.~Weber}\altaffiliation[Present address: ]{Siemens, Munich, Germany}

\columbia

\begin{abstract} 
The dependence of the light and charge yield of liquid xenon on the applied electric field and recoil energy is important for dark matter detectors using liquid xenon time projections chambers. Few measurements have been made of this field dependence at recoil energies less than 10\,keV. In this paper we present results of such measurements using a specialized detector. Recoil energies are determined via the Compton coincidence technique at four drift fields relevant for liquid xenon dark matter detectors: 0.19, 0.48, 1.02, and 2.32\,kV/cm. Mean recoil energies down to 1\,keV were measured with unprecedented precision. We find that the charge and light yield are anti-correlated above $\sim$3\,keV, and that the field dependence becomes negligible below $\sim$6\,keV. However, below 3\,keV we find a charge yield significantly higher than expectation and a reconstructed energy deviating from linearity.

\end{abstract}
\pacs{
 29.40.Mc, 
 61.25.Bi, 
 78.70.-g, 
 95.35.+d, 
}
\keywords{Liquid Xenon, Charge Yield, Light Yield, Dark Matter}
\maketitle

\section{Introduction}
The next generation of dark matter detectors using liquid xenon (LXe) as the detection medium \cite{Aprile_xenon1t,lz} will probe lower interaction cross sections than ever before. One key to success for these experiments is an improved understanding of low-energy interactions in LXe. The interaction cross sections between weakly interacting massive particles (WIMPs) and Standard Model particles are generally expected to increase exponentially with decreasing interaction energy. However, low-energy interactions in LXe, below a few keV, are not well-understood theoretically, and measurements of the fundamental properties of LXe at low energies that allow the interaction energy scale to be precisely defined are lacking. 

For instance, systematic uncertainties on the scintillation-based energy scale have limited the search for axions \cite{Aprile:axion} and for annual modulation \cite{Aprile:dcpaper,Aprile:acpaper} with XENON100 data.
These systematic uncertainties could be improved by using a combined energy scale (CES), that is by using both the prompt scintillation ($S1$) and ionization ($S2$) signals, provided that the anti-correlation of $S1$ and $S2$ in LXe at low energies is well characterized. Charge yield measurements of electronic recoils (ERs) are currently most needed. 
Lack of such measurements is the primary reason a CES was not adopted for studies of ERs in XENON100. Next generation detectors using a CES will benefit from these measurements.

Measurements of low-energy ERs from monochromatic $\gamma$ rays are of particular interest for dark matter searches using LXe, as their dominant electromagnetic backgrounds in the outer layer of the detector come from such Compton scatters.

Direct measurements have been made of the light yield of LXe due to ERs from monochromatic $\gamma$ rays as a function of the recoil energy down to low energies \cite{Baudis_2013,Aprile_2012a}. These yields were derived from measurements of $S1$ using single-phase LXe detectors operated at either zero drift field or at a single applied electric field. The lowest mean energy previously measured for the light yield due to ERs is 1.5\,keV \cite{Baudis_2013}.

Even fewer measurements have been made of the charge yield of LXe due to ERs, derived from the $S2$ signal, since collecting the ionization electrons typically requires two-phase operation, entailing more sophisticated detectors. 
At the current time, no references have been found for dedicated measurements of the charge yield of ERs at low energies using Compton scattering of external monochromatic $\gamma$-ray sources, the method employed for this work.
The charge yield of ERs has been measured by several groups as a function of electric field using mono-chromatic $\gamma$ rays, but only at relatively high energies, e.g. 122, 511, and 570\,keV. See Ref.\,\cite{Szydagis_2011} for a review of ER light and charge measurements. 

The light and charge yield from the decay of $^{83m}$Kr at 9.4\,keV and 32.1\,keV has been measured \citep{Baudis_2013,Aprile_2012a,Manalaysay_2010}, though the decay products consist primarily of several internal conversion and Auger electrons, and thus are not directly comparable to measurements with monochromatic $\gamma$ rays. Additionally, the charge and light yield from the decay of tritium ($^3$H) has been measured at two drift fields \citep{Akerib_2016}. However, as tritium is a single-beta emitter with a Q-value of 18.6\,keV, the resulting energy spectrum is continuous, and thus Monte Carlo (MC) matching is required for the recoil energy determination. Similarly, continuous spectra from the decay of $^{137}$Cs have been measured and an accompanying MC used to determine the charge and light yield in LXe at various drift fields \citep{Lin_2015}.

In this paper we present results of measurements of the light and charge yield of LXe due to low-energy ERs as a function of applied electric field and deposited energy. The Compton coincidence technique is used to determine the amount of energy deposited in an ER in the LXe. The technique, introduced in Ref.\,\cite{Valentine_1994,Rooney_1996} and further improved in Ref.\,\cite{Choong_2008a,Choong_2008b}, is the same technique used for the measurement of the light yield of ERs described in detail in Ref.\,\cite{Aprile_2012a}. For this work we employ the same method, but now use a two-phase time projection chamber (TPC), thus allowing both the charge and light yield to be measured simultaneously. Science data presented here were taken during two periods, referred to as runs I and II. 

\section{Experimental Setup}
The principle of the Compton coincidence technique (CCT) is to irradiate the primary detector with high-energy monochromatic $\gamma$ rays, and to look for events in a second detector in prompt coincidence  corresponding to the full-energy deposition of the Compton-scattered $\gamma$ rays. 
Since the initial energy is known and the final energy is measured by the second detector, the energy lost in the primary detector can be determined. By placing the second detector at various angles relative to the incoming $\gamma$ rays, different recoil energy ranges can be probed in the primary detector. 

\begin{figure}
\centering
\includegraphics[width=0.7\columnwidth]{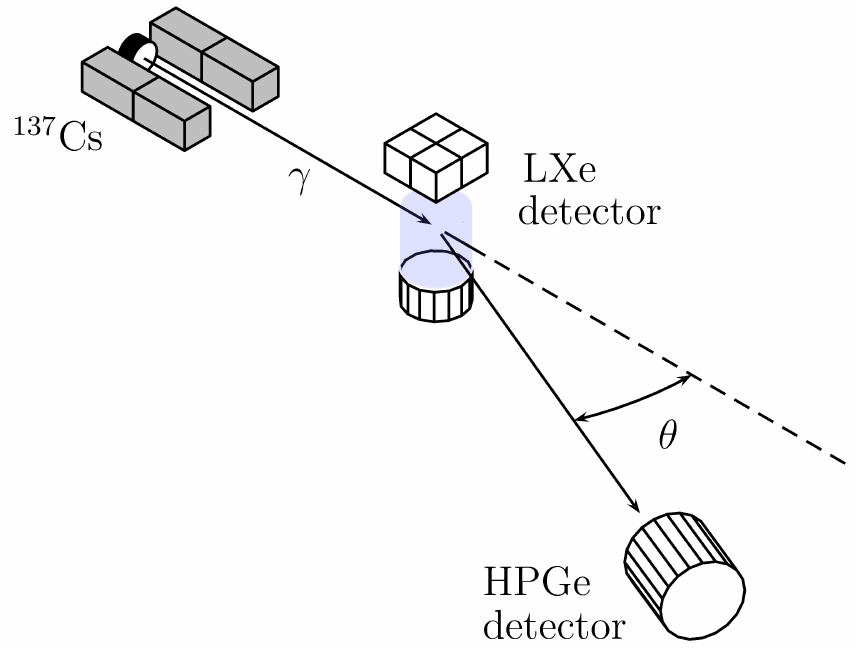}
\vspace*{-0.2cm}
\caption{\label{method_schematic} 
Schematic of the experimental setup during Compton coincidence measurements. Monochromatic $\gamma$ rays from the decay of $^{137}$Cs Compton scatter in the LXe detector, depositing energy which results in electronic recoils. The energy of the $\gamma$ rays scattering at an angle $\theta$ is measured by a HPGe detector. \vspace{-0.12in}}
\end{figure}

We measure recoils from incoming 661.7\,keV $\gamma$ rays emitted from a 123\,$\mu$Ci $^{137}$Cs source, and measure the scattered $\gamma$ rays with a high-purity germanium (HPGe) detector, as shown schematically in \figref{method_schematic}. 

\begin{figure}
\centering
\includegraphics[width=1.\columnwidth]{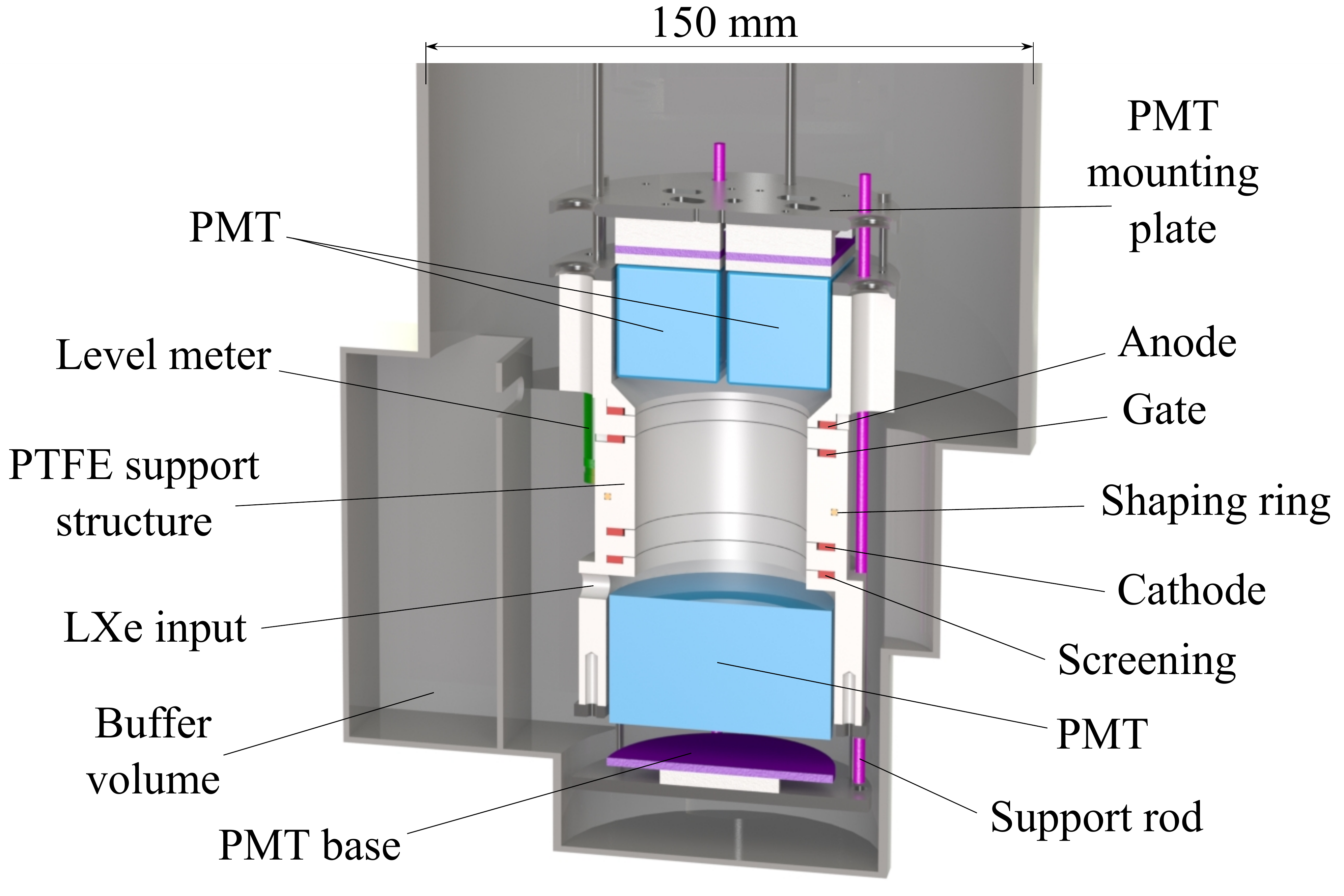}
\vspace*{-0.5cm}
\caption{\label{detector_schematic} 
Schematic drawing of the neriX detector.
}
\end{figure}

The primary detector used for this measurement is a two-phase LXe TPC designed to measure both the light and charge yield of nuclear and electronic recoils in LXe, and is thus dubbed neriX. The neriX detector is the latest iteration of such detectors designed and built at Columbia University for measuring the fundamental properties of LXe. As shown in \figref{detector_schematic}, the neriX detector consists of five photomultiplier tubes (PMTs) surrounding a sensitive volume enclosed by a series of grids supported in a polytetrafluoroethylene (PTFE) frame. Four 1" square PMTs (Hamamatsu R8520-406-M4-SEL), each with four channels, form a square top array of 16 channels in the gaseous Xe above the sensitive volume. One 2" diameter circular PMT (Hamamatsu R6041-406 SEL) views the LXe volume from below. All PMTs were selected for having high quantum efficiency (QE), $>$35\% at the Xe scintillation wavelength, 178\,nm. During coincidence measurements, the top PMT array is used for event vertex reconstruction and the bottom PMT is used for the yield determination.

The PTFE support pieces are stackable and compressed vertically between two stainless steel plates via polyether ether ketone (PEEK) support rods with stainless steel springs. The springs ensure that the stack remains compressed and relatively light tight when at LXe temperature. The nearly seamless PTFE inner surface surrounding the sensitive volume ensures optimal light collection efficiency. The bottom and top PMTs are mounted to separate stainless steel plates that can be translated vertically with respect to the TPC stack. 

To produce the electric fields in the liquid and gaseous Xe, high voltage (HV) is applied to hexagonal mesh grids with 3\,mm pitch, etched from stainless steel foil. The anode, gate, and cathode grids are 125\,$\mu$m thick, and the screening grid directly above the bottom PMT is 25\,$\mu$m thick, ensuring high optical transparency and good electric field uniformity. The gate and screening grids are electrically grounded, and negative HV is applied to the PMTs and the cathode, while $+$4.5\,kV is applied to the anode. A 56.5\,mm diameter ring, made from 1.65\,mm diameter Cu wire is fixed 7\,mm above the cathode and coaxial with the TPC. Two 500\,M$\Omega$ resistors connected to the ring divide the voltage between the cathode and gate grids, improving the linearity of the electron drift field in the sensitive volume. 

The sensitive volume, that is the volume from which both light and charge signals from the LXe can be measured, is a cylinder 23.4\,mm tall with a 43\,mm diameter while at typical LXe temperature, between the gate and cathode grids. The volume contains a LXe target mass of approximately 97\,g. 

A custom vacuum cryostat vessel made from 1.5\,mm thick stainless steel was designed to minimize the amount of LXe and other material outside of the neriX TPC and provide the LXe level-setting mechanism. In addition, the amount of PTFE around the sensitive volume was minimized in order to reduce the likelihood that $\gamma$ rays deposit energy outside of the sensitive volume. A total of $\sim$2.2\,kg of LXe are used to fill the detector vessel.  

The neriX cryogenic and purification system was adapted from that used for the measurements in Ref.\,\cite{Aprile_2012a,Plante_2011}. An emergency recovery vessel and rupture disk were added to capture the Xe gas in the case of a loss of cooling power or insulation vacuum. Circulation of the Xe gas through the purification loop was achieved using a diaphragm pump (KNF 143 SN.12E).

The secondary detector is a p-type coaxial HPGe detector (Ortec GEM40-76) with a crystal of 63.7\,mm diameter and 59.6\,mm length. The typical full width at half maximum (FWHM) energy resolution at 1.33\,MeV is 1.81\,keV, with a typical Peak-to-Compton ratio of 71:1. 

During coincidence measurements, the $^{137}$Cs source was placed 17.6\,cm horizontally from the TPC center, and the HPGe detector was placed such that the center of the detector endcap was 14.6\,cm horizontally from the TPC center. The positions were measured to within 1\,mm using aluminum rules and an auto-leveling laser. The relative angle between the $^{137}$Cs and HPGe detector was fixed at either 0$^{\circ}$ or 25$^{\circ}$ during science data taking. Different configurations of the source and HPGe detector positions were used for the two runs in order to study the impact of the LXe buffer volume. 
The bottom PMT used during run I experienced a deterioration of gain, and was thus exchanged in between the runs for another PMT of the same type and comparable QE.

The data acquisition system for neriX is similar to that used in Ref.\,\cite{Aprile_2012a,Plante_2011}. The raw data from neriX is amplified x10 (Phillips 776) and one copy is digitized by three 14-bit flash ADCs (CAEN V1724) at 100\,MS/s with 40\,MHz bandwidth, and stored to disk. The other copy is fed into the trigger logic system. 

The HPGe detector provides two copies of the pre-amplified HPGe signal. One copy is inverted and passed through a leading edge discriminator
thus producing the HPGe trigger. The other copy is amplified with a pulse-shaping amplifier using a coarse gain of x25 and a 6\,$\mu$s shaping time, digitized by the same CAEN system, and then recorded to disk. During measurements at 25$^{\circ}$, the raw signals from the PMTs are not amplified, so as to mitigate the saturation of the digitizer due to large $S2$ signals from higher energy interactions.

Signals from the PMT channels pass through a leading edge discriminator
which subsequently produces the $S1$ hardware trigger.
The discriminator is updating, producing a 10\,ns pulse whenever an input signal is above the threshold, resulting in variable-length trigger pulses. 
Logical operations ensure that an $S2$ trigger is subsequently produced only for pulses longer than 100\,ns.
Measurements of the $S2$ trigger efficiency were made using back-to-back 511\,keV $\gamma$ rays from the decay of $^{22}$Na in prompt coincidence with a NaI detector. The average trigger efficiency for $S2$ with 10$^3$\,PE is $\sim$75\%, and is thus near 100\% for most recoil energies measured. Fits of the observed $S2$ spectra are corrected for this efficiency loss. The impact on the attributed charge yield is at the percent level for 1\,keV recoils, and diminishes with increasing recoil energy.

The hardware trigger for coincidence events is formed by requiring that an $S2$ trigger come within $20-30$\,$\mu$s after an HPGe trigger, depending on the drift field being measured.
The coincidence trigger and $S1$ trigger are recorded to disk.  
A stricter time coincidence requirement between the signals from the two detectors is imposed in post-processing, as described in Section\,\ref{data_sel}.

\section{Calibration}
\label{cal}
\subsection{LXe detector calibration}
\label{lxe_cal}

Calibration measurements of the PMTs single PE gain were taken biweekly using a blue LED. Typical gains were $(5-7)\cdot 10^5$\,electrons/PE. Regular calibrations of neriX with $\gamma$ rays from 
$^{137}$Cs (662\,keV) were taken to monitor the TPC performance. 
The typical measured light yields for $^{137}$Cs during runs I and II are given in Table\,\ref{typical_ly}.

\begin{table}
\caption{Typical measured light yield of neriX for the full-absorption peak of
$^{137}$Cs (661.7~keV) at various drift fields, V$_{\text{D}}$, during runs I and II. Uncertainties represent statistical and systematic uncertainties added in quadrature.}
\label{typical_ly}
\begin{center}
    \begin{ruledtabular}
	\begin{tabular}{c c c} 
	V$_{\text{D}}$ [kV/cm] & \multicolumn{2}{c}{Average light yield [PE/keV]}\\
	\hline
	&  run I & run II \\
	0.19 & 2.84 $\pm$ 0.04 & 2.79 $\pm$ 0.08\\
	0.48 & 2.57 $\pm$ 0.03 & 2.41 $\pm$ 0.04\\
	1.02 & 2.27 $\pm$ 0.03 & 2.18 $\pm$ 0.06\\
	2.32 & 2.11 $\pm$ 0.04 & 1.94 $\pm$ 0.06\\
	\end{tabular} 
	\end{ruledtabular}
\end{center}
\vspace*{-0.3cm}
\end{table}

The room temperature QE of the bottom PMT used in neriX was measured by the manufacturer to be 39.6\% and 37.6\% for the different PMTs used in run I and II, respectively. 

For each run, the liquid level was set at the position that optimized the $S2$ energy resolution. This resulted is an increase in the average level of 310\,$\mu$m between the runs. The level was observed to be stable to within $\pm 50$\,$\mu$m during run I and $\pm 20$\,$\mu$m during run II. 

The neriX detector is capable of measuring single electron signals, thus allowing for the absolute gain of extracted ionization electrons to be measured directly. In this work, the single electron gain, $G$, is defined as the number of PE detected by the bottom PMT for each extracted electron. This gain was measured approximately once per week, or whenever the drift field was changed, using $\gamma$ rays from $^{137}$Cs and following the method described in Ref.\,\cite{Aprile_2014a}. The average measured $G$ for each cathode voltage is summarized in Table\,\ref{nerix_g2_ave_table}. Typical gains were $(16-17)$\,PE/electron. The systematic difference in $G$ between the runs is accounted for by the difference in QE of the PMTs and the change in liquid level. The efficiency of electron extraction, $\eta$, was estimated assuming the relation in Eq.\,\ref{anticorr} as discussed in Section\,\ref{yield}. The loss of electrons due to attachment to impurities in the LXe was found to be negligible.

\begin{table}
\caption{Average single electron gain, $G$, in neriX detector for the bottom PMT only for the different drift fields, V$_{\text{D}}$, measured over the course of runs I and II. Uncertainties represent statistical and systematic uncertainties added in quadrature.}
\label{nerix_g2_ave_table}
\begin{center}
    \begin{ruledtabular}
	\begin{tabular}{c c c} 
	V$_{\text{D}}$ [kV/cm] & \multicolumn{2}{c}{Average $G$ [PE/electron]}\\	
	\hline
	&  run I & run II \\
	0.19 & 16.1 $\pm$ 0.5 & 17.3 $\pm$ 0.6\\
	0.48 & 16.9 $\pm$ 0.5 & 17.1 $\pm$ 0.8\\
	1.02 & 16.8 $\pm$ 0.5 & 17.6 $\pm$ 0.7\\
	2.32 & 16.6 $\pm$ 0.5 & 17.2 $\pm$ 0.7\\
	\end{tabular} 
	\end{ruledtabular}
\end{center}
\vspace*{-0.3cm}
\end{table}

\subsection{HPGe detector calibration}
\label{hpge_cal}
The HPGe detector performance was monitored on a weekly basis and was calibrated using $\gamma$ rays from $^{57}$Co (122.1\,keV and 136.5\,keV), $^{137}$Cs (661.7\,keV) and $^{22}$Na (1275\,keV). The detector response was found to be linear with an energy resolution well described by $\sigma_{HPGe}(E) = \alpha + \beta \cdot \sqrt{E}$, where $E$ is the $\gamma$-ray energy, and the values of $\alpha$ and $\beta$ are determined from the fit. The resulting energy calibration was found to be stable to within 2\% of the mean over the course of runs I and II. A weighted average of the results of all calibrations measurements was taken to determine the energy calibration used for the final analysis of each run. 
The average FWHM energy resolution at 662\,keV and 552\,keV was determined to be 1.4\,keV and 1.3\,keV, respectively. 
Hence, when using the CCT method to determine the recoil energy in LXe, the $1\sigma$ uncertainty on the inferred energy varies between 0.6\,keV for $E_r=0.5$\,keV and 0.5\,keV for $E_r=110$\,keV.

\section{Compton Coincidence Measurements}
Coincidence measurements were made over the course of runs I and II for different angular configurations and at different cathode voltages (drift fields). In run I, measurements were made at 0$^{\circ}$ and 25$^{\circ}$ for cathode voltages of $-0.345$, $-1.054$, $-2.356$, and $-5.550$\,kV. In run II, measurements were made at 0$^{\circ}$ for the same voltages, and at 25$^{\circ}$ for cathode voltages of $-2.356$ and $-5.550$\,kV only. The total live time of coincidence measurements was 41.1\,days and 28.0\,days in runs I and II, respectively. During each run, science data were recorded nearly 24 hours per day, with brief, regular pauses for the calibration measurements described in Section\,\ref{cal}. 

\subsection{Electric field simulation and fiducial volume}
\label{efield_sim}
The electric field in the neriX detector was simulated using COMSOL Multiphysics\textsuperscript{\textregistered} Suite. An accurate model of the relevant detector geometry and materials was employed to calculate the spatial uniformity of the drift field for various cathode, anode, and PMT voltages. Simulations in both 2D and 3D were performed as a consistency check. For the cathode voltages employed for the final coincidence measurements, $-0.345$, $-1.054$, $-2.356$, and $-5.500$\,kV, the mean drift fields as estimated by both 2D and 3D simulations of the detector and cryostat are 0.19 $\pm$ 0.03, 0.48 $\pm$ 0.05, 1.02 $\pm$ 0.12, and 2.32 $\pm$ 0.28\,kV/cm, respectively. For these cases, the anode voltage is fixed to be $+$4.5\,kV, and the gate and screening grids held at 0\,V. See Table\,\ref{vd_table} for a summary.

These simulations show that the drift field is very uniform near the center of the detector, but deviates non-linearly near the grids and detector walls. The fiducial volume (FV) for the final analysis was chosen so as to maximize the accepted number of coincidence events while ensuring an acceptable spatial variation of the drift field. To this end, the FV was defined with boundaries 1\,mm below the gate grid and above the cathode grid, and 0.5\,mm radially inward from the inner surface of the PTFE walls. 
The systematic uncertainties on the drift field reported here are a volumetric average of the variation of the simulated field over the extent of the FV due to the non-homogeneity and anisotropy of the detector geometry.

\subsection{Monte Carlo simulation}
\label{mc}
The expected performance of the neriX detector during coincidence measurements was simulated using Geant4. The Geant4 geometry includes a detailed model of the neriX detector, cryostat, and outer support structures, as well as the HPGe detector. The simulation was used to estimate the expected effect of energy loss outside of the TPC sensitive volume due to multiple scattering. 

These simulations show that for recoils with energies $<20$\,keV, on average less than 45\,eV is lost outside of the FV. Thus, the expected systematic shift in the energy attributed to coincidence events is only $\sim$5\% for 1\,keV recoils, and less for higher-energy recoils. For all recoil energies considered in this analysis, this shift is negligible compared to the energy resolution of the HPGe detector. In addition, these simulations show that the average energy lost outside of the FV for the different configurations of the source and HPGe detector positions used in run I and II is negligibly different ($<2$\,eV). Hence, the data from these runs can be combined.

\subsection{Data processing and selection}
\label{data_sel}
The recorded data consists of 17 PMT waveforms, the amplified HPGe signal, and a multiplexed trigger signal. The data is processed using the same peak-finding software as XENON100 \cite{Aprile_2012b}. For each scintillation signal identified by the software, the position, area, height, and other basic parameters are computed. The trigger signal is de-multiplexed and the number of $S1$ and $S2$-HPGe coincidence triggers and their positions in the trace are extracted.
 
The event vertex in the x-y plane is found using a neural network model. 
The uncertainty on the position reconstruction is estimated to be less than 3\,mm, as the anode grid, which has a 3\,mm pitch, is resolved. The depth of events is found using the difference in time between the $S1$ and $S2$ signals and the measured field-dependent drift velocity, $v_d$. The average $v_d$ in neriX is measured directly using photoionization of the grids, following the method described in Ref.\,\cite{Aprile_2014a}, but using ionization from $S2$ rather than from $S1$. The $v_d$ measurements for the various drift fields are summarized in Table\,\ref{vd_table}.

\begin{table}
\caption{Measured average electron drift velocities in LXe at various cathode voltages, V$_{\text{C}}$, and drift fields, V$_{\text{D}}$, used for final science measurements.}
\label{vd_table}
\begin{center}
    \begin{ruledtabular}
	\begin{tabular}{ccccc}
	V$_{\text{C}}$ [kV] & $-0.345$ & $-1.054$ & $-2.356$ & $-5.550$ \\
	\hline
	V$_{\text{D}}$ [kV/cm] & 0.19 & 0.48 & 1.02 & 2.32 \\ 
	$\pm 1\sigma$ [kV/cm] & 0.03 & 0.05 & 0.12 & 0.28 \\ 
	\hline 
	$v_d$ [mm/$\mu$s], run I & 1.51 & 1.72 & 1.96 & 2.21 \\  
	$v_d$ [mm/$\mu$s], run II & 1.54 & 1.75 & 1.97 & 2.22 \\ 
	\end{tabular} 
	\end{ruledtabular}
\end{center}
\vspace*{-0.3cm}
\end{table}

Events are selected by requiring that they pass a certain set of basic quality cuts. For $S1$ spectra, they must have a single valid $S1$ and $S2$ with an event vertex within the FV, defined
 in Section\,\ref{efield_sim}. They must also pass a timing cut, which requires that the $S1$ peak be identified in the trace within 0.5\,$\mu$s before a HPGe trigger.
The observed distribution of the relative time between the $S1$ and HPGe trigger is strongly peaked near $\sim$0.25\,$\mu$s, and increasing the time window beyond 0.5\,$\mu$s has negligible impact on the results. The size of the $S1$ signals is corrected according to their depth, and the size of all signals are scaled by the measured PMT gains. For $S2$ spectra, the requirement of an $S1$ is dropped so as to avoid any possible impact from a loss of $S1$ detection efficiency, and the timing of the HPGe trigger is used to determine the depth of events. The measured $S2$ trigger efficiency is taken into account. The $S2$ signals are not corrected for electron loss by attachment to impurities in the LXe, as such charge loss is found to be negligible. The valid coincidence events are then divided into recoil energy ranges by selecting energy subranges of the corresponding HPGe signals.

\subsection{Measured distributions}
\begin{figure*}
	\begin{center}
		\includegraphics[width=0.95\textwidth]{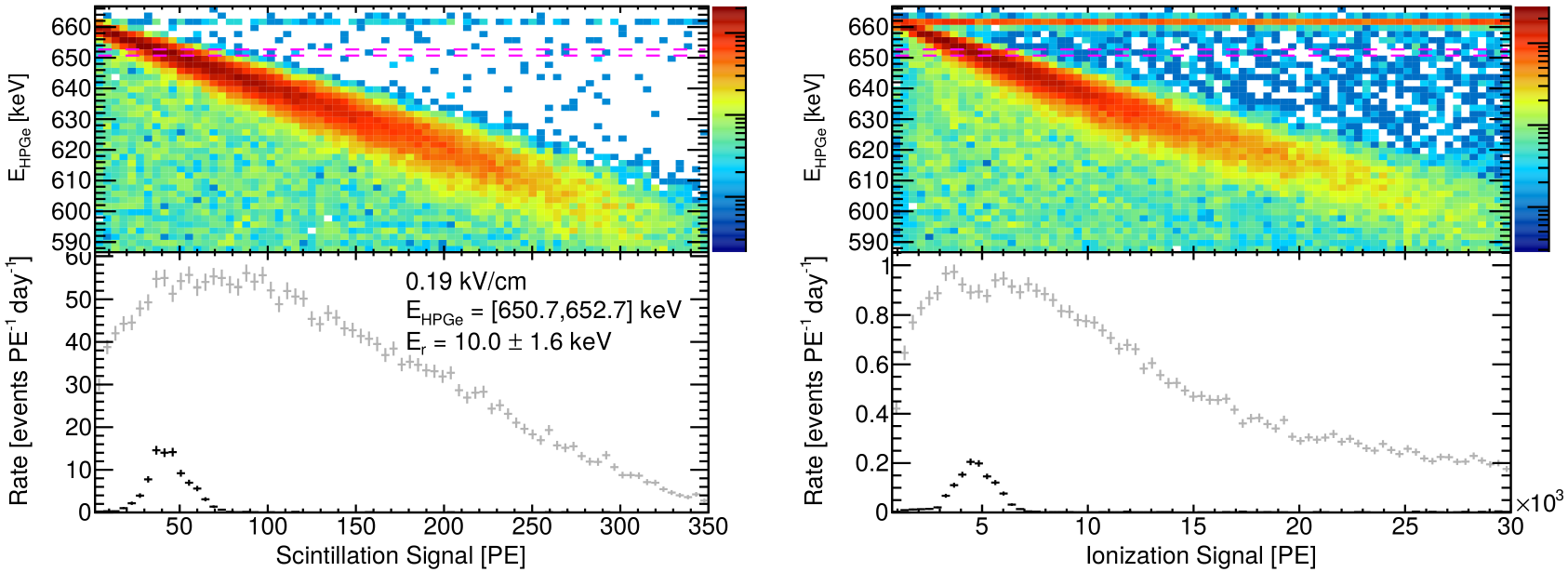}
		\caption{Distribution of measured HPGe signal (vertical axis in top panels) versus the scintillation signal (left), and ionization signal (right) for coincidence data taken at 0\,$^{\circ}$ and 0.19\,kV/cm drift field during run I. The selected energy subrange in the HPGe, ($650.7-652.7$)\,keV, is designated by the dashed lines in the top panels, and corresponds to ERs in neriX with $E_r= 10.0 \pm 1.6$\,keV. The light gray histograms show the corresponding full light and charge distributions in neriX, while the black histograms show the distributions for the selected energy subrange.}\vspace{-0.15in}
		\label{hpge_dist}
	\end{center}
\end{figure*}

An example of the measured HPGe and neriX distributions for valid coincidence events is shown in the top panels of \figref{hpge_dist}. The main distributions show a clear correlation between the energy deposited in the HPGe and LXe, as expected, and correspond to events where all of the energy is deposited in the two detectors. The approximately flat distribution of events below this curve corresponds to events where not all the energy is deposited, e.g. to events where the $\gamma$ ray Compton scatters from the HPGe detector, depositing only a fraction of its energy. The timing cut largely removes events above this curve, corresponding to accidental coincidence events. Events with full-energy deposition in the HPGe (662\,keV) in accidental coincidence with events in neriX are also clearly visible. The energy of these events and spread therein are in good agreement with the dedicated HPGe calibrations described in Section\,\ref{hpge_cal}, and validate the stability of the calibration throughout the course of science data taking. 

Histograms of the corresponding light ($S1$) and charge ($S2$) distributions in neriX are shown in the bottom panels of \figref{hpge_dist}. 
The widths of the HPGe energy subranges were chosen so that there are at least several hundred events per energy bin, thus ensuring sufficient statistics for all spectral fits, and better than 1$\%$ statistical uncertainty for most bins. An example of this energy selection is shown in Fig.\,\ref{hpge_dist}. Approximately equal statistics were taken for data at 0\,$^{\circ}$ and 0.19, 1.02, and 2.32\,kV/cm drift fields, allowing for 2\,keV wide HPGe energy bins. Roughly double the statistics were taken at 0\,$^{\circ}$ and 0.48\,kV/cm drift field, allowing for 1\,keV wide HPGe energy bins. For data at 25\,$^{\circ}$, 4\,keV wide HPGe energy bins are employed for data taken at all fields.

\begin{figure*}[h]
	\begin{center}
		\includegraphics[width=0.86\textwidth]{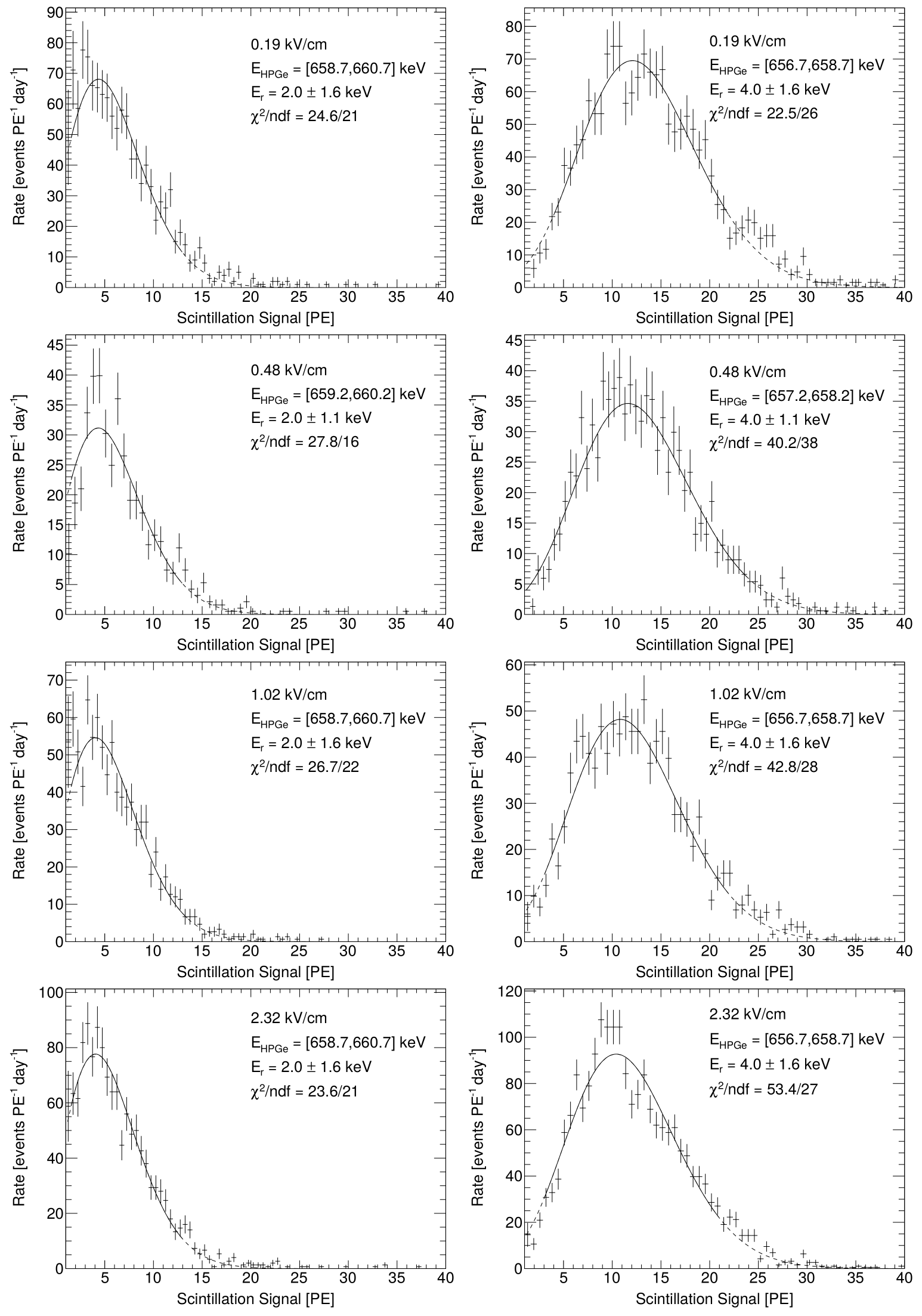}
		\vspace*{-0.3cm}
		\caption{Distribution of scintillation signal, $S1$, at various drift fields for 2\,keV (left) and 4\,keV (right) electronic recoils in neriX during run II coincidence measurements. Smooth curves show fit according to Eq.\,\ref{fit_func} (dotted line), and fit range (solid line).}
		\label{s1_dist_lowe}
	\end{center}
\end{figure*}

\begin{figure*}[h]
	\begin{center}
		\includegraphics[width=0.86\textwidth]{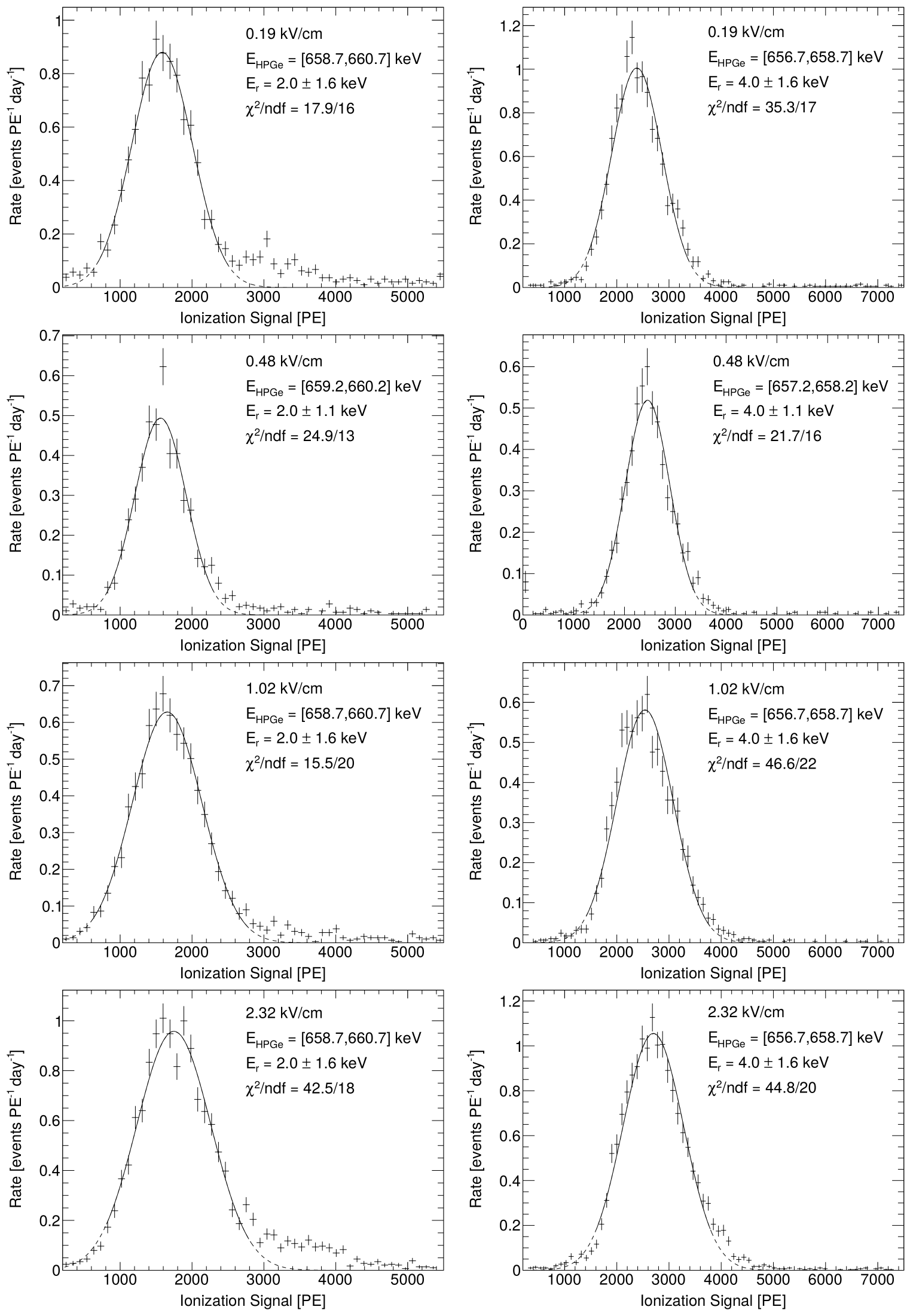}
		\vspace*{-0.3cm}
		\caption{Distribution of ionization signal, $S2$, at various drift fields for 2\,keV (left) and 4\,keV (right) electronic recoils in neriX during run II coincidence measurements. Smooth curves show Gaussian fit (dotted line), and fit range (solid line).}
		\label{s2_dist_lowe}
	\end{center}
\end{figure*}

\subsection{Yield determination}
\label{yield}

The light and charge yields were determined by selecting recoil energy subranges in the HPGe detector of $1-2$\,keV width, depending on the data type, fitting the corresponding $S1$ and $S2$ distributions to find the average light and charge signal size for each range, and then dividing by the respective recoil energy in the center of the range, $E_{r}$. The recoil energy is calculated as $E_{r} = 661.7- E_{HPGe}$ (in keV). The light and charge yields were extracted for recoils with mean energies down to 1\,keV at 0.48\,kV/cm drift field, and down to 2\,keV at 0.19, 1.02, and 2.32\,kV/cm drift fields.

The $S1$ spectra for low-energy subranges, $\leq 10$\,keV, are fit by a sum of Poisson functions, modeling the emission and collection process of scintillation light, each convolved with a corresponding Gaussian function, modeling the smearing of the signal due to the amplification process. Specifically, the fit function takes the form
\begin{equation}
\label{fit_func}
f(x) = \sum_{n} Ae^{-\mu}\mu^n \cdot (n!)^{-1} \cdot (\sqrt{2\pi n}\sigma)^{-1} \cdot e^{-\frac{(x-n)^2}{2n\sigma^2}}
\end{equation} 
where $A$ is the amplitude, $\mu$ is the mean $S1$ response, $\langle S1 \rangle$, and $\sigma$ is the effective resolution per PE. 
The sum runs over the range $n=[1,100]$\,PE, and $f(x)$ is proportional to the probability for the reconstructed signal, $S1$, 
to be of size $x$ if $\mu$\,PE are detected on average over the considered energy interval.
For each fit, $A$ and $\mu$ are free parameters and $\sigma$ is fixed to the average best-fit value over the energy range of 1.32. The resulting shape is slightly asymmetric for small $S1$ signals, but becomes increasingly symmetric for $E_r$ greater than a few keV.

The observed $S1$ spectra have a rate falling off at low PE like a binomial, as expected due to the PMT QE and the geometrical limitation of the absolute light collection efficiency. The low-energy analysis threshold for $S1$ was hence chosen to be 1.5\,PE, corresponding to a theoretical absolute detection efficiency of 80\%. No correction for this efficiency loss was included in the spectral fits.

For $S2$ signals (with means typically $>10^3$\,PE at minimum) and for $S1$ signals for $E_r$ $>$10\,keV, a simple Gaussian function is used to fit and determine $\langle S1 \rangle$ and $\langle S2 \rangle$. Examples of the spectra fitting at 2 and 4\,keV are shown in Fig.\,\ref{s1_dist_lowe}-\ref{s2_dist_lowe}.  

Saturation of the measured $S2$ signals due to saturation of the digitizer was observed for some data taken at 0\,$^{\circ}$. Above a certain recoil energy threshold, $\sim 24-40$\,keV depending on the cathode voltage, this saturation increased with recoil energy.
The maximum energy at which the charge yield is reported here for data taken at 0\,$^{\circ}$ was chosen accordingly, so as to remove the affected data. 

\begin{figure*}[ht]
\centering
\includegraphics[width=\textwidth]{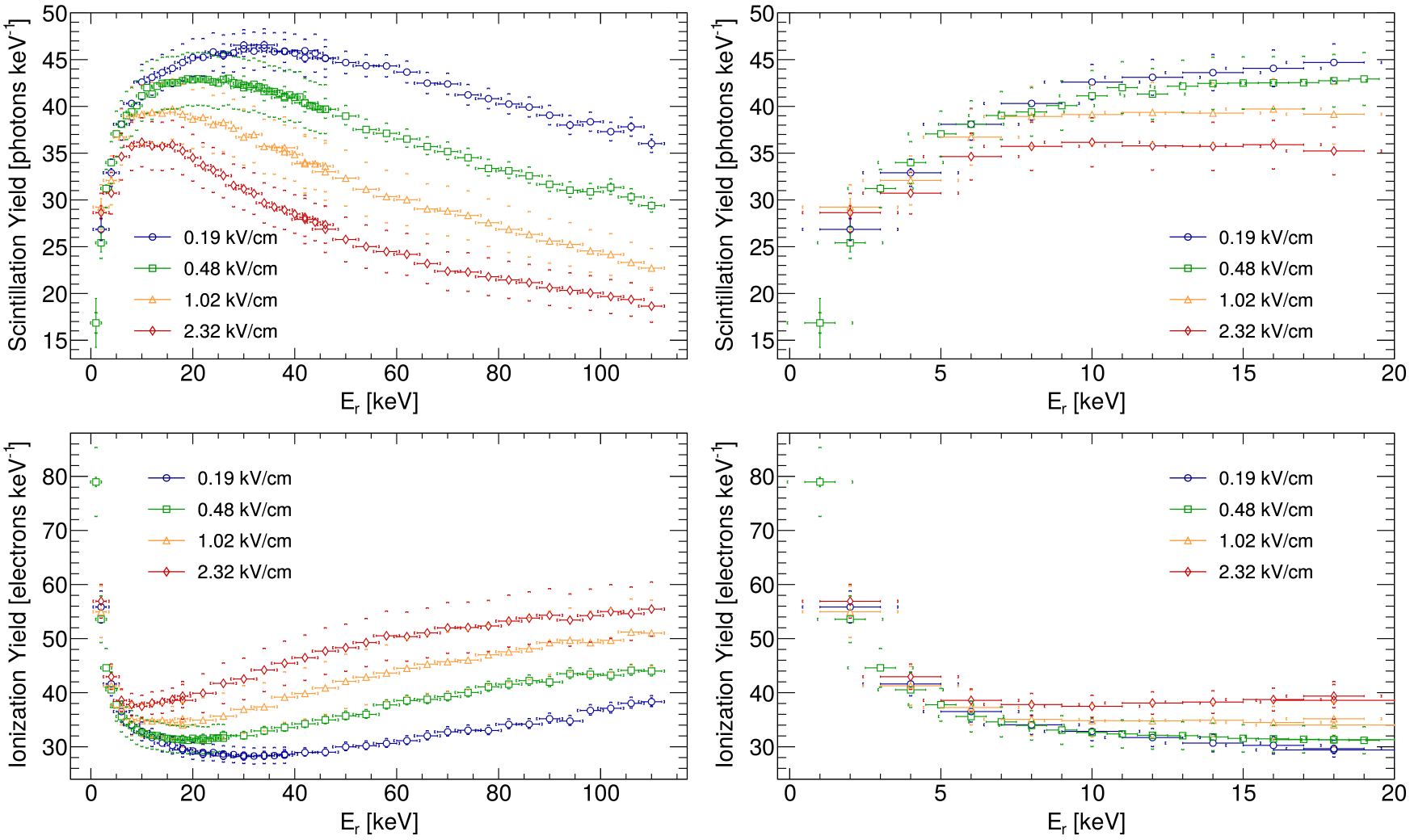}
\vspace*{-0.6cm}
\caption{\label{ly_ab} 
Light yield (top panels) and charge yield (bottom panels) as a function of recoil energy and drift field for ERs in neriX, with low-energy zooms (right panels). Weighted fit of results from runs I and II for data taken at 0$^{\circ}$ and 25$^{\circ}$. Statistical and recoil-energy dependent uncertainties on yields shown as solid vertical lines. Systematic uncertainties on yields shown as open vertical square brackets. Selected recoil energy range shown as solid horizontal lines, with open brackets showing expected extent due to finite HPGe energy resolution. Corresponding point values and uncertainties are given in Tables\,\ref{345V_table}-\ref{5500V_table}.\vspace{-0.12in}
}
\end{figure*}

No saturation effects are apparent in the data taken at 25\,$^{\circ}$ due to the bypassing of the x10 amplifier. However, a systematic rise in the light yield with decreasing recoil energy is evident below a certain energy threshold. No opposite trend is observed for the corresponding $S2$ data, and no similar trend is observed in the $S1$ data with the same recoil energies for data taken at 0\,$^{\circ}$. Above this threshold, the yields from measurements at 0\,$^{\circ}$ and 25\,$^{\circ}$ are in very good agreement, as evident in Fig.\,\ref{ly_ab}. The threshold above which this effect was not apparent was between $25-65$\,keV, depending on the drift field. The nature of this systematic effect is not currently understood. Affected data were removed from the results presented here.

The total numbers of scintillation photons and ionization electrons emitted per unit recoil energy, after electron-ion recombination has taken place, are referred to as the light and charge yields, respectively. These absolute yields, in units of photons/keV and electrons/keV, are determined from the measured yields, in PE/keV, by scaling by the absolute gains of the $S1$ and $S2$ signals, referred to herein as $g1$ and $g2$. Specifically, for this analysis $g1$ and $g2$ are defined as the average number of PE detected by the bottom PMT for each scintillation photon and ionization electron escaping electron-ion recombination, respectively. In general, the value of $g1$ depends on the light collection efficiency of the detector as well as the collection efficiency and QE of the PMT. The value of $g2$ depends on the gas pressure and electric field strength near and above the liquid surface, and hence primarily on the anode voltage and liquid level height. 

Anti-correlation between the expectation values of $S1$ and $S2$ is assumed, such that
\begin{equation}
\label{anticorr}
E_{r}/W = \langle S1 \rangle/g1 + \langle S2 \rangle/g2
\end{equation}
where $W$ is the average energy required to produce either one scintillation photon or ionization electron. We assume an average value for $W$ of 13.7 $\pm$ 0.4\,eV \cite{Szydagis_2011}. Values of $\langle S1 \rangle$ and $\langle S2 \rangle$ are extracted for each energy subrange at each drift field measured. A plot of $\langle S1 \rangle/E_{r}$ versus $\langle S2 \rangle/E_{r}$ results in a straight line which is subsequently fit assuming Eq.\,\ref{anticorr} to determine $g1$ and $g2$ for each data type, as described below.  

Direct measurements of $G$ for each cathode voltage measured, described in Sec\,\ref{lxe_cal} and summarized in Table\,\ref{nerix_g2_ave_table}, are related to $g2$ by the efficiency for extracting ionization electrons, $\eta$, such that $g2 = \eta \cdot G$. 

A global fit of the measured yields using a Markov Chain Monte Carlo (MCMC) is performed \cite{mcmc}, taking into account the measured values of $W$, $G$, and the PMT gains in order to determine the best-fit values of $W$, $g1$, $\eta$, and $G$. Gaussian priors are assumed for $W$, $G$, and the PMT gains, while flat priors are assumed for $g1$ and $\eta$. The energy range considered for this analysis is $E_r > 10.5$\,keV, i.e. well above detector threshold and in an energy range where Eq.\,\ref{anticorr} has been validated experimentally. 

The results of this analysis are summarized in Table\,\ref{ancor_table}. The best-fit $W$ values are lower than the accepted value, but in good agreement given uncertainties. The difference in $g1$ values between the runs is in agreement with the expectation, given the different QE of the PMTs for the two runs. The difference in $\eta$ values between the runs is attributed to the different liquid level in those runs, while the field dependence is attributed to a change of the field strength near the liquid/gas interface due to leakage of the gate grid. In general it is found that correlations between the fit parameters are non-negligible, and that the derived probability distributions for $\eta$ are not Gaussian due to the physical constraint that $\eta <1$. Nonetheless, comparable fit results and uncertainties (apart from uncertainties on $\eta$ for mean values with $\eta \approx 1$) are obtained using a traditional $\chi^2$ analysis and propagating uncertainties using the full covariance matrix of the fit.

\begin{table}
\caption{Best-fit values from the anti-correlation analysis described in text.}
\label{ancor_table}
\begin{center}
    \begin{ruledtabular}
	\begin{tabular}{cccc}
	Parameter & V$_{\text{D}}$ [kV/cm] & run I & run II \\
	\hline
	$W$ [eV] & \multirow{2}{*}{All} & 13.4 $\pm$ 0.3 & 13.4 $\pm$ 0.4\\
	$g1$ [PE/photon ]& & 0.105 $\pm$ 0.003 & 0.099 $\pm$ 0.005 \\ 
	\hline
	\multirow{4}{*}{$\eta$} & 0.19 & 0.89$^{+0.04}_{-0.04}$ & 0.78$^{+0.07}_{-0.06}$ \\
	& 0.48 & 0.95$^{+0.03}_{-0.04}$ & 0.82$^{+0.09}_{-0.09}$ \\
	& 1.02 & 0.96$^{+0.02}_{-0.03}$ & 0.90$^{+0.06}_{-0.09}$ \\
	& 2.32 & 0.95$^{+0.03}_{-0.04}$ & 0.96$^{+0.03}_{-0.05}$ \\	
	\hline
	\multirow{4}{*}{$G$ [PE/electron]} & 0.19 & 16.2 $\pm$ 0.4 & 17.3 $\pm$ 0.6 \\
	& 0.48 & 16.9 $\pm$ 0.4 & 17.1 $\pm$ 0.8 \\
	& 1.02 & 16.9 $\pm$ 0.3 & 17.6 $\pm$ 0.7 \\	
	& 2.32 & 16.9 $\pm$ 0.4 & 17.6 $\pm$ 0.7 \\	
	\end{tabular} 
	\end{ruledtabular}
\end{center}
\vspace*{-0.3cm}
\end{table}

\begin{figure}
\centering
\includegraphics[width=0.98\columnwidth]{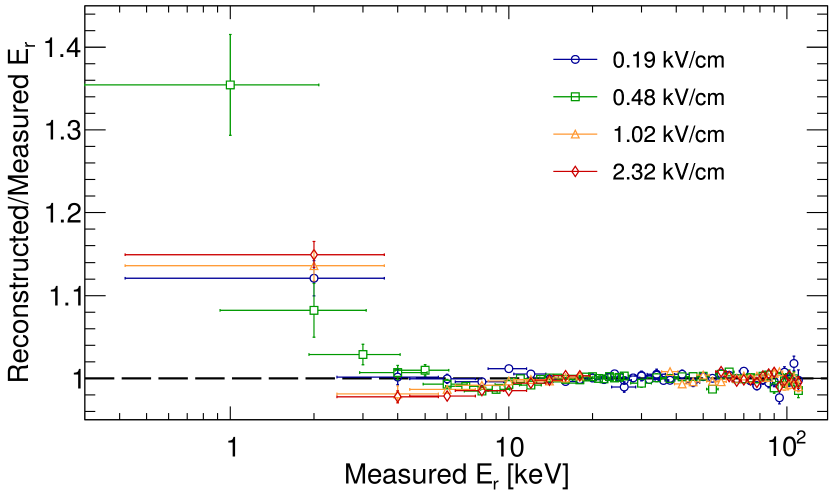}
\vspace*{-0.3cm}
\caption{\label{en_recon} 
Relative deviation of the recoil energy as reconstructed using Eq.\,\ref{anticorr} from the recoil energy as measured by the HPGe detector, shown versus the measured recoil energy. The expectation of no deviation is shown as a dashed line.
\vspace{-0.12in}
}
\end{figure}

The posterior distributions from the MCMC are sampled in order to determine the yields in photons/keV and electrons/keV, as well as the reconstructed mean recoil energies for each run. The weighted average of the yields and reconstructed energies from runs I and II is computed using the statistical and systematic uncertainties added in quadrature as the weights. The resulting average yields are shown in Fig.\,\ref{ly_ab}. The relative deviation of the average reconstructed recoil energies from the measured recoil energies is shown in Fig.\,\ref{en_recon}.

\subsection{Yield uncertainties}
\label{uncert}

Due to the finite energy resolution of the HPGe detector, the range of actual recoil energies in each energy bin is broader than the nominal width. Thus, while different energy bins do not share events, they are expected to overlap somewhat in terms of the actual recoil energy of events in the bins. The selected recoil energy range for each bin is shown as solid horizontal error bars in Fig.\,\ref{ly_ab}. The estimated range of actual recoil energies, that is the selected energy range plus the $1 \sigma$ HPGe energy resolution, is shown as open horizontal brackets in Fig.\,\ref{ly_ab} and as solid horizontal error bars in Fig.\,\ref{en_recon}. 

Since the yields are determined by dividing by the energy of each bin center, a shift in the average energy of events in each bin from this center energy would result in a systematic shift the attributed mean yield. However, it is observed that the average energy of events for each bin is within 10$\%$ of the bin center for the lowest-energy recoil of each data type (1 or 2\,keV), and within 1$\%$ of the bin center for all other bins. In all cases, the resulting shift in the attributed mean energy is negligible compared to the bin width and energy resolution of the HPGe detector.

The systematic effect of the binning definition and fit range of the measured $S1$ and $S2$ distributions on the yield determination was studied and found to impact the $S1$ fits in the lowest $1-2$ bins for data taken at 0$^{\circ}$ only. The average relative sizes of these effects were 16\% and 9\%, respectively, for the lowest energy bin, and 4\% and 5\%, respectively, for the following bin. By the third bin, these effects were negligible. The corresponding systematic uncertainties in the light yield, which are recoil-energy dependent, are added in quadrature with the statistical uncertainties of the respective bins and are shown as solid vertical bars around each point in Fig.\,\ref{ly_ab} and reported in Tables\,\ref{345V_table}-\ref{5500V_table}.

The variation of the actual yield and differential scattering rate over the width of each energy bin can also result in a systematic shift of the attributed mean yield. Given the observed variation of the measured yield between consecutive bins (taking into account the uncertainties mentioned above) and assuming a flat differential scattering rate, we estimate this possible shift to be at most $\sim$9\% for 1\,keV recoils, $1-6$\% for recoils between 2\,keV and 6\,keV, and $<1$\% for recoils $>6$\,keV. The actual size of such a systematic shift is difficult to quantify since the actual yields are unknown. This possible shift is not included in the uncertainties reported herein.

The uncertainties on the PMT gain, $G$, $g1$, and $\eta$ affect all points of a given data type uniformly, and are herein referred to as systematic uncertainties. The correlations between these parameters are probed by the MCMC of the global fit of the reconstructed energy, which is used to propagate the uncertainties to the final yields and reconstructed recoil energy for each run and data type. The resulting total systematic uncertainties are shown as vertical open brackets around each point in Fig.\,\ref{ly_ab} and are reported in Tables\,\ref{345V_table}-\ref{5500V_table}. In Fig.\,\ref{en_recon}, the vertical error bars show the statistical and systematic errors added in quadrature. 

\section{Results}

The final results of the light and charge yield determinations are shown in Fig.\,\ref{ly_ab} and summarized in Tables\,\ref{345V_table}-\ref{5500V_table}.
We observe that the light yield increases with recoil energy, attaining a maximum between $10-35$\,keV depending on the field, and then decreases. A similar, but opposite, trend is observed for the charge yield. The general dependence of the yield on the drift field is significant and is as expected for a charge recombination process, showing a decreasing light yield and increasing charge yield with increasing drift field. We also find that the drift field dependence of the yields diminishes with decreasing recoil energy and becomes negligible, given experimental uncertainties, below $\sim 6$\,keV.

The reconstructed energies, Fig.\,\ref{en_recon}, show good agreement with the anti-correlation relationship as formulated in Eq.\,\ref{anticorr} for recoil energies above 3\,keV. Below 3\,keV a significant deviation from expectation is observed, increasing from an average of $+12\%$ at 2\,keV to $+35\%$ at 1\,keV.

\begin{figure}
\centering
\includegraphics[width=\columnwidth]{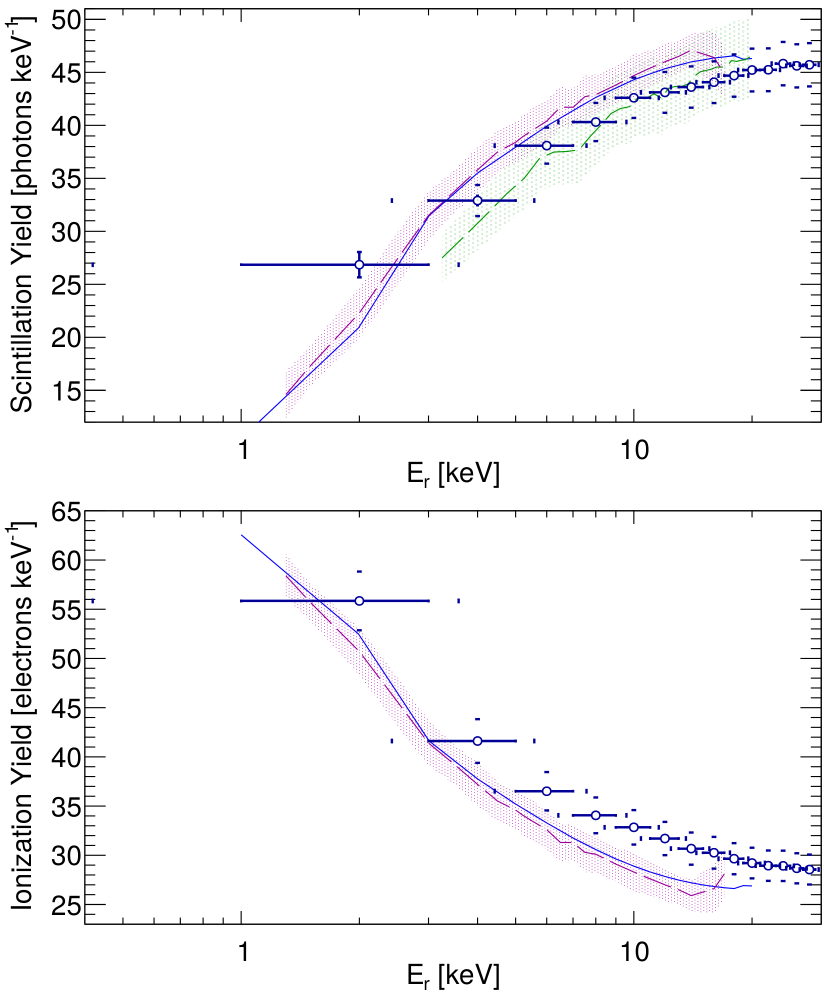}
\vspace*{-0.5cm}
\caption{\label{comp} 
Comparison plot of LXe light yield (top) and charge yield (bottom) from ERs with comparable electric field strengths. The neriX results taken at 0.19\,$\pm$\,0.03\,kV/cm (blue circles)
 are shown together with measurements from \citep{Akerib_2016} taken at 0.18\,kV/cm (magenta band and dashed line), measurements from \citep{Lin_2015} taken at 0.236\,kV/cm (green band and dashed line), and calculations from \citep{Akerib_2016,nest_v1} at 0.19\,V/cm (blue line).
\vspace{-0.12in}
}
\end{figure}

See Fig.\,\ref{comp} for a comparison of our results taken at 0.19\,kV/cm with recent measurements and predictions at comparable drift fields.   

\begin{figure}
\centering
\includegraphics[width=0.85\columnwidth]{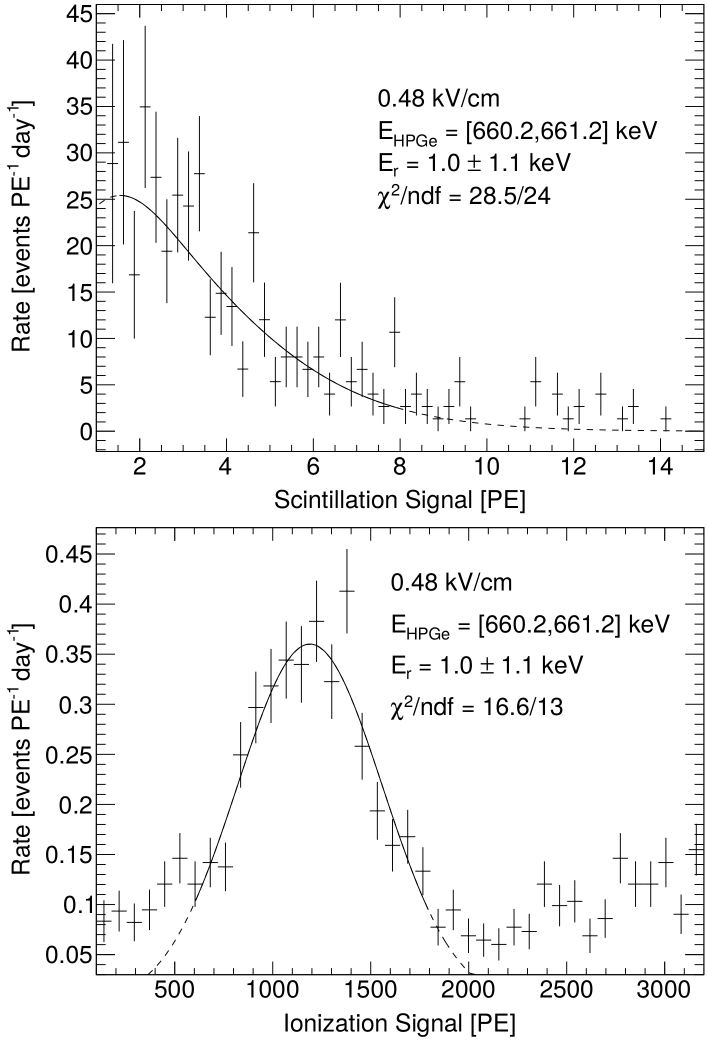}
\vspace{-0.2cm}
\caption{\label{s2_1kev} Distribution of scintillation signal (top panel) and ionization signal (bottom panel) for 1\,keV recoils at 0.48\,kV/cm drift field during run II.
\vspace{-0.12in}
}
\end{figure}

\section{Discussion}
We find that our results at 0.19\,kV/cm are generally in good agreement with recent measurements at comparable fields using continuous recoil energy bands. The measured light yield agrees very well with that in \citep{Lin_2015}, while both are systematically lower than that in \citep{Akerib_2016} by $\sim 1\sigma$ on average. The measured charge yield is systematically higher than that in \citep{Akerib_2016} by $\sim 1-2\sigma$.

Our results are in similarly good agreement with the NEST model predictions \citep{Akerib_2016,nest_v1} at low energies $>2$\,keV. The agreement for the light and charge yields at 0.48, 1.02, and 2.32\,kV/cm is very good, though for data at 0.19\,kV/cm the measured field dependence is less pronounced than that predicted by NEST. For 2\,keV recoils, however, we observe a slightly higher light yield than predicted by NEST at all measured fields, and for 1\,keV recoils at 0.48\,kV/cm, the observed charge yield is significantly higher.

Fig.\,\ref{en_recon} shows that our data supports the use of a combined energy scale as defined in Eq.\,\ref{anticorr} for determining ER energies in LXe dark matter detectors for recoil energies above $\sim3$\,keV. However, for recoil energies below $\sim3$\,keV, a significant deviation from the expectation is observed. Even assuming a light yield of zero at 1\,keV, the measured charge yield exceeds what should be possible given the accepted, recoil-energy independent average $W$ value. Assuming an ionization potential of 11.7\,eV for LXe, the maximum number of quanta that can be produced for 1\,keV recoils is $\sim85.5$, which is greater than the measured charge yield, but in agreement given experimental uncertainties.

A threshold effect on the $S2$ signal could plausibly cause an apparent rise in the charge yield. However, the measured $S2$ trigger efficiency is incorporated in the analysis. Furthermore, the observed $S2$ spectra for 1\,keV recoils, see Fig.\,\ref{s2_dist_lowe} and \ref{s2_1kev}, show no evidence of significant asymmetry. In addition, reducing the energy bin width to 0.5\,keV for the data taken at 0.48\,kV/cm drift field results in a nearly identical charge yield at 1\,keV.
Thus, it is unlikely that the high charge yield at 1\,keV is due to a rate or binning effect. 

A threshold effect on the $S1$ signal, e.g. a loss of the $S1$ peak identification efficiency of the raw data processing software as well as the decrease in $S1$ detection efficiency due to the expected statistical limitations mentioned in Sec\,\ref{yield}, could also potentially cause such an apparent rise in the charge yield. However, dropping the $S1$ requirement for the $S2$ spectra ensures that such effects can have no impact. 

Thus, we conclude that the observed rise in the charge yield is likely not artificial, and may instead indicate that the combined energy scale as described in Eq.\,\ref{anticorr} breaks down at very low energies, namely that the average $W$ value may be energy dependent.

\begin{table*}
\caption{Weighted fit of measured light and charge yields in neriX detector during runs I and II for data taken at a mean drift field of 0.19 $\pm$ 0.03\,kV/cm, assuming $g1$, $\eta$, and $G$ values stated in text.}
\label{345V_table}
\subfloat{\begin{tabular}{cccc}
	\hline \hline
	\multicolumn{4}{c}{Source-HPGe Angle: 0$^\circ$} \\
	\multicolumn{4}{c}{Light Yield Relative Sys. Uncertainty: 4.5\,$\%$} \\
	\multicolumn{4}{c}{Charge Yield Relative Sys. Uncertainty: 5.4\,$\%$} \\
	\hline 
	$E_{r}$ Range & HPGe Res. & Light Yield & Charge Yield \\
	{[}keV] & [keV] & [photons/keV] & [electrons/keV] \\  
	\hline
	$1.0-3.0$ & 0.6 & 26.9 $\pm$ 1.2 & 55.8 $\pm$ 0.4 \\
	$3.0-5.0$ & 0.6 & 32.9 $\pm$ 0.5 & 41.6 $\pm$ 0.2 \\
	$5.0-7.0$ & 0.6 & 38.1 $\pm$ 0.2 & 36.5 $\pm$ 0.1 \\
	$7.0-9.0$ & 0.6 & 40.3 $\pm$ 0.2 & 34.1 $\pm$ 0.1 \\
	$9.0-11.0$ & 0.6 & 42.6 $\pm$ 0.2 & 32.8 $\pm$ 0.1 \\
	$11.0-13.0$ & 0.6 & 43.1 $\pm$ 0.2 & 31.7 $\pm$ 0.1 \\
	$13.0-15.0$ & 0.6 & 43.6 $\pm$ 0.1 & 30.7 $\pm$ 0.1 \\
	$15.0-17.0$ & 0.6 & 44.1 $\pm$ 0.1 & 30.3 $\pm$ 0.1 \\
	$17.0-19.0$ & 0.6 & 44.7 $\pm$ 0.1 & 29.7 $\pm$ 0.1 \\
	$19.0-21.0$ & 0.6 & 45.2 $\pm$ 0.1 & 29.2 $\pm$ 0.1 \\
	$21.0-23.0$ & 0.6 & 45.2 $\pm$ 0.1 & 28.9 $\pm$ 0.1 \\
	$23.0-25.0$ & 0.6 & 45.8 $\pm$ 0.1 & 28.9 $\pm$ 0.1 \\
	$25.0-27.0$ & 0.6 & 45.6 $\pm$ 0.1 & 28.7 $\pm$ 0.1 \\
	$27.0-29.0$ & 0.6 & 45.7 $\pm$ 0.1 & 28.6 $\pm$ 0.1 \\
	$29.0-31.0$ & 0.6 & 46.1 $\pm$ 0.1 & 28.4 $\pm$ 0.1 \\
	$31.0-33.0$ & 0.6 & 45.9 $\pm$ 0.1 & 28.3 $\pm$ 0.1 \\
	$33.0-35.0$ & 0.6 & 46.2 $\pm$ 0.1 & 28.4 $\pm$ 0.1 \\
	$35.0-37.0$ & 0.6 & 45.8 $\pm$ 0.1 & 28.4 $\pm$ 0.1 \\
	$37.0-39.0$ & 0.6 & 45.9 $\pm$ 0.1 & 28.4 $\pm$ 0.1 \\
	$39.0-41.0$ & 0.6 & 45.7 $\pm$ 0.1 & -- \\
	$41.0-43.0$ & 0.6 & 45.2 $\pm$ 0.1 & -- \\
	$43.0-45.0$ & 0.6 & 45.7 $\pm$ 0.1 & -- \\
	$45.0-47.0$ & 0.6 & 45.1 $\pm$ 0.1 & -- \\
	\hline \hline
	\end{tabular}} 
\qquad
\subfloat{\begin{tabular}{cccc}
	\hline \hline
	\multicolumn{4}{c}{Source-HPGe Angle: 25$^\circ$} \\
	\multicolumn{4}{c}{Light Yield Relative Sys. Uncertainty: 2.7\,$\%$} \\
	\multicolumn{4}{c}{Charge Yield Relative Sys. Uncertainty: 2.7\,$\%$} \\
	\hline 
	$E_{r}$ Range & HPGe Res. & Light Yield & Charge Yield \\
	{[}keV] & [keV] & [photons/keV] & [electrons/keV] \\   
	\hline
	$16.0-20.0$ & 0.6 & -- & 29.4 $\pm$ 0.2 \\
	$20.0-24.0$ & 0.6 & -- & 28.7 $\pm$ 0.2 \\
	$24.0-28.0$ & 0.6 & 45.4 $\pm$ 0.4 & 28.3 $\pm$ 0.2 \\
	$28.0-32.0$ & 0.6 & 46.5 $\pm$ 0.4 & 28.2 $\pm$ 0.2 \\
	$32.0-36.0$ & 0.6 & 46.6 $\pm$ 0.4 & 28.3 $\pm$ 0.2 \\
	$36.0-40.0$ & 0.6 & 45.9 $\pm$ 0.3 & 28.6 $\pm$ 0.2 \\
	$40.0-44.0$ & 0.6 & 45.9 $\pm$ 0.3 & 28.9 $\pm$ 0.2 \\
	$44.0-48.0$ & 0.6 & 45.2 $\pm$ 0.3 & 29.0 $\pm$ 0.2 \\
	$48.0-52.0$ & 0.6 & 44.7 $\pm$ 0.3 & 30.0 $\pm$ 0.2 \\
	$52.0-56.0$ & 0.6 & 44.3 $\pm$ 0.2 & 30.2 $\pm$ 0.2 \\
	$56.0-60.0$ & 0.6 & 44.3 $\pm$ 0.3 & 30.6 $\pm$ 0.2 \\
	$60.0-64.0$ & 0.6 & 43.7 $\pm$ 0.2 & 31.1 $\pm$ 0.2 \\
	$64.0-68.0$ & 0.6 & 42.5 $\pm$ 0.2 & 32.0 $\pm$ 0.2 \\
	$68.0-72.0$ & 0.6 & 42.4 $\pm$ 0.2 & 32.8 $\pm$ 0.2 \\
	$72.0-76.0$ & 0.6 & 41.2 $\pm$ 0.2 & 33.1 $\pm$ 0.2 \\
	$76.0-80.0$ & 0.6 & 40.8 $\pm$ 0.2 & 33.0 $\pm$ 0.3 \\
	$80.0-84.0$ & 0.6 & 40.3 $\pm$ 0.3 & 34.1 $\pm$ 0.3 \\
	$84.0-88.0$ & 0.6 & 39.9 $\pm$ 0.3 & 34.2 $\pm$ 0.3 \\
	$88.0-92.0$ & 0.6 & 39.1 $\pm$ 0.3 & 35.2 $\pm$ 0.4 \\
	$92.0-96.0$ & 0.6 & 38.0 $\pm$ 0.4 & 34.8 $\pm$ 0.4 \\
	$96.0-100.0$ & 0.6 & 38.4 $\pm$ 0.3 & 36.7 $\pm$ 0.4 \\
	$100.0-104.0$ & 0.6 & 37.3 $\pm$ 0.3 & 37.1 $\pm$ 0.5 \\
	$104.0-108.0$ & 0.6 & 37.8 $\pm$ 0.4 & 38.0 $\pm$ 0.5 \\
	$108.0-112.0$ & 0.5 & 36.0 $\pm$ 0.5 & 38.3 $\pm$ 0.7 \\
	\hline \hline
	\end{tabular}} 
\end{table*}

\begin{table*}
\caption{Weighted fit of measured light and charge yields in neriX detector during runs I and II for data taken at a mean drift field of 0.48 $\pm$ 0.05\,kV/cm, assuming $g1$, $\eta$, and $G$ values stated in text.}
\label{1054V_table}
\subfloat{\begin{tabular}{cccc}
    \hline \hline
	\multicolumn{4}{c}{Source-HPGe Angle: 0$^\circ$} \\
	\multicolumn{4}{c}{Light Yield Relative Sys. Uncertainty: 6.6\,$\%$} \\
	\multicolumn{4}{c}{Charge Yield Relative Sys. Uncertainty: 8.1\,$\%$} \\
	\hline 
	$E_{r}$ Range & HPGe Res. & Light Yield & Charge Yield \\
	{[}keV] & [keV] & [photons/keV] & [electrons/keV] \\  
	\hline
	$0.5-1.5$ & 0.6 & 16.9 $\pm$ 2.6 & 79.0 $\pm$ 0.9 \\
	$1.5-2.5$ & 0.6 & 25.4 $\pm$ 1.0 & 53.6 $\pm$ 0.3 \\
	$2.5-3.5$ & 0.6 & 31.2 $\pm$ 0.4 & 44.6 $\pm$ 0.2 \\
	$3.5-4.5$ & 0.6 & 34.0 $\pm$ 0.3 & 40.5 $\pm$ 0.1 \\
	$4.5-5.5$ & 0.6 & 37.1 $\pm$ 0.3 & 37.8 $\pm$ 0.1 \\
	$5.5-6.5$ & 0.6 & 38.1 $\pm$ 0.2 & 35.6 $\pm$ 0.1 \\
	$6.5-7.5$ & 0.6 & 39.0 $\pm$ 0.2 & 34.6 $\pm$ 0.1 \\
	$7.5-8.5$ & 0.6 & 39.4 $\pm$ 0.2 & 33.9 $\pm$ 0.1 \\
	$8.5-9.5$ & 0.6 & 40.1 $\pm$ 0.2 & 33.1 $\pm$ 0.1 \\
	$9.5-10.5$ & 0.6 & 41.1 $\pm$ 0.2 & 32.6 $\pm$ 0.1 \\
	$10.5-11.5$ & 0.6 & 42.0 $\pm$ 0.2 & 32.4 $\pm$ 0.1 \\
	$11.5-12.5$ & 0.6 & 41.3 $\pm$ 0.2 & 32.1 $\pm$ 0.1 \\
	$12.5-13.5$ & 0.6 & 42.2 $\pm$ 0.1 & 32.0 $\pm$ 0.1 \\
	$13.5-14.5$ & 0.6 & 42.5 $\pm$ 0.1 & 31.8 $\pm$ 0.1 \\
	$14.5-15.5$ & 0.6 & 42.5 $\pm$ 0.1 & 31.5 $\pm$ 0.1 \\
	$15.5-16.5$ & 0.6 & 42.5 $\pm$ 0.1 & 31.5 $\pm$ 0.1 \\
	$16.5-17.5$ & 0.6 & 42.5 $\pm$ 0.1 & 31.4 $\pm$ 0.1 \\
	$17.5-18.5$ & 0.6 & 42.7 $\pm$ 0.1 & 31.3 $\pm$ 0.1 \\
	$18.5-19.5$ & 0.6 & 42.9 $\pm$ 0.1 & 31.2 $\pm$ 0.1 \\
	$19.5-20.5$ & 0.6 & 42.8 $\pm$ 0.1 & 31.6 $\pm$ 0.1 \\
	$20.5-21.5$ & 0.6 & 42.9 $\pm$ 0.1 & 31.3 $\pm$ 0.1 \\
	$21.5-22.5$ & 0.6 & 42.9 $\pm$ 0.1 & 31.2 $\pm$ 0.1 \\
	$22.5-23.5$ & 0.6 & 42.9 $\pm$ 0.1 & 31.2 $\pm$ 0.1 \\
	$23.5-24.5$ & 0.6 & 42.6 $\pm$ 0.1 & 31.6 $\pm$ 0.1 \\
	$24.5-25.5$ & 0.6 & 42.5 $\pm$ 0.1 & 31.6 $\pm$ 0.1 \\
	$25.5-26.5$ & 0.6 & 42.9 $\pm$ 0.1 & 31.6 $\pm$ 0.1 \\
	$26.5-27.5$ & 0.6 & 43.0 $\pm$ 0.1 & -- \\
	$27.5-28.5$ & 0.6 & 42.5 $\pm$ 0.1 & -- \\
	$28.5-29.5$ & 0.6 & 42.5 $\pm$ 0.1 & -- \\
	$29.5-30.5$ & 0.6 & 42.3 $\pm$ 0.1 & -- \\
	$30.5-31.5$ & 0.6 & 42.2 $\pm$ 0.1 & -- \\
	$31.5-32.5$ & 0.6 & 42.4 $\pm$ 0.1 & -- \\
	$32.5-33.5$ & 0.6 & 42.1 $\pm$ 0.1 & -- \\
	$33.5-34.5$ & 0.6 & 42.0 $\pm$ 0.1 & -- \\
	$34.5-35.5$ & 0.6 & 41.6 $\pm$ 0.1 & -- \\
	$35.5-36.5$ & 0.6 & 41.6 $\pm$ 0.1 & -- \\
	$36.5-37.5$ & 0.6 & 41.4 $\pm$ 0.1 & -- \\
	$37.5-38.5$ & 0.6 & 41.1 $\pm$ 0.1 & -- \\
	$38.5-39.5$ & 0.6 & 41.2 $\pm$ 0.1 & -- \\
	$39.5-40.5$ & 0.6 & 41.0 $\pm$ 0.1 & -- \\
	$40.5-41.5$ & 0.6 & 41.0 $\pm$ 0.1 & -- \\
	$41.5-42.5$ & 0.6 & 40.3 $\pm$ 0.1 & -- \\
	$42.5-43.5$ & 0.6 & 40.1 $\pm$ 0.1 & -- \\
	$43.5-44.5$ & 0.6 & 40.0 $\pm$ 0.1 & -- \\
	$44.5-45.5$ & 0.6 & 39.7 $\pm$ 0.1 & -- \\
	$45.5-46.5$ & 0.6 & 39.7 $\pm$ 0.1 & -- \\
	\hline \hline
	\end{tabular}}
\qquad
\subfloat{\begin{tabular}{cccc}
    \hline \hline
	\multicolumn{4}{c}{Source-HPGe Angle: 25$^\circ$} \\
	\multicolumn{4}{c}{Light Yield Relative Sys. Uncertainty: 2.7\,$\%$} \\
	\multicolumn{4}{c}{Charge Yield Relative Sys. Uncertainty: 2.3\,$\%$} \\
	\hline 
	$E_{r}$ Range & HPGe Res. & Light Yield & Charge Yield \\
	{[}keV] & [keV] & [photons/keV] & [electrons/keV] \\  
	\hline
	$16.0-20.0$ & 0.6 & -- & 31.3 $\pm$ 0.2 \\
	$20.0-24.0$ & 0.6 & -- & 31.5 $\pm$ 0.2 \\
	$24.0-28.0$ & 0.6 & -- & 32.1 $\pm$ 0.2 \\
	$28.0-32.0$ & 0.6 & 42.0 $\pm$ 0.3 & 32.0 $\pm$ 0.2 \\
	$32.0-36.0$ & 0.6 & 41.6 $\pm$ 0.3 & 32.9 $\pm$ 0.2 \\
	$36.0-40.0$ & 0.6 & 40.9 $\pm$ 0.3 & 33.5 $\pm$ 0.2 \\
	$40.0-44.0$ & 0.6 & 40.4 $\pm$ 0.2 & 34.2 $\pm$ 0.2 \\
	$44.0-48.0$ & 0.6 & 39.7 $\pm$ 0.2 & 35.0 $\pm$ 0.2 \\
	$48.0-52.0$ & 0.6 & 39.0 $\pm$ 0.2 & 35.7 $\pm$ 0.2 \\
	$52.0-56.0$ & 0.6 & 37.5 $\pm$ 0.2 & 36.0 $\pm$ 0.2 \\
	$56.0-60.0$ & 0.6 & 37.1 $\pm$ 0.2 & 37.7 $\pm$ 0.2 \\
	$60.0-64.0$ & 0.6 & 36.5 $\pm$ 0.2 & 38.6 $\pm$ 0.2 \\
	$64.0-68.0$ & 0.6 & 35.7 $\pm$ 0.2 & 38.8 $\pm$ 0.2 \\
	$68.0-72.0$ & 0.6 & 35.2 $\pm$ 0.2 & 39.3 $\pm$ 0.2 \\
	$72.0-76.0$ & 0.6 & 34.5 $\pm$ 0.2 & 40.0 $\pm$ 0.3 \\
	$76.0-80.0$ & 0.6 & 33.4 $\pm$ 0.2 & 41.1 $\pm$ 0.3 \\
	$80.0-84.0$ & 0.6 & 33.1 $\pm$ 0.3 & 41.6 $\pm$ 0.3 \\
	$84.0-88.0$ & 0.6 & 32.6 $\pm$ 0.3 & 42.2 $\pm$ 0.3 \\
	$88.0-92.0$ & 0.6 & 31.7 $\pm$ 0.3 & 42.0 $\pm$ 0.4 \\
	$92.0-96.0$ & 0.6 & 31.0 $\pm$ 0.3 & 43.5 $\pm$ 0.4 \\
	$96.0-100.0$ & 0.6 & 30.9 $\pm$ 0.3 & 43.4 $\pm$ 0.4 \\
	$100.0-104.0$ & 0.6 & 31.3 $\pm$ 0.5 & 43.2 $\pm$ 0.5 \\
	$104.0-108.0$ & 0.6 & 30.3 $\pm$ 0.4 & 44.1 $\pm$ 0.4 \\
	$108.0-112.0$ & 0.5 & 29.4 $\pm$ 0.4 & 44.0 $\pm$ 0.5 \\
	\hline \hline
	\end{tabular}}
\end{table*}

\begin{table*}
\caption{Weighted fit of measured light and charge yields in neriX detector during runs I and II for data taken at a mean drift field of 1.02 $\pm$ 0.12\,kV/cm, assuming $g1$, $\eta$, and $G$ values stated in text.}
\label{2356V_table}
\subfloat{\begin{tabular}{cccc}
	\hline \hline
	\multicolumn{4}{c}{Source-HPGe Angle: 0$^\circ$} \\
	\multicolumn{4}{c}{Light Yield Relative Sys. Uncertainty: 8.2\,$\%$} \\
	\multicolumn{4}{c}{Charge Yield Relative Sys. Uncertainty: 8.7\,$\%$} \\
	\hline 
	$E_{r}$ Range & HPGe Res. & Light Yield & Charge Yield \\
	{[}keV] & [keV] & [photons/keV] & [electrons/keV] \\  
	\hline
	$1.0-3.0$ & 0.6 & 29.2 $\pm$ 0.9 & 55.0 $\pm$ 0.3 \\
	$3.0-5.0$ & 0.6 & 32.1 $\pm$ 0.5 & 41.3 $\pm$ 0.2 \\
	$5.0-7.0$ & 0.6 & 36.7 $\pm$ 0.2 & 37.3 $\pm$ 0.2 \\
	$7.0-9.0$ & 0.6 & 38.9 $\pm$ 0.2 & 35.1 $\pm$ 0.1 \\
	$9.0-11.0$ & 0.6 & 39.1 $\pm$ 0.2 & 34.8 $\pm$ 0.1 \\
	$11.0-13.0$ & 0.6 & 39.4 $\pm$ 0.2 & 34.8 $\pm$ 0.1 \\
	$13.0-15.0$ & 0.6 & 39.3 $\pm$ 0.2 & 34.9 $\pm$ 0.1 \\
	$15.0-17.0$ & 0.6 & 39.7 $\pm$ 0.2 & 34.5 $\pm$ 0.1 \\
	$17.0-19.0$ & 0.6 & 39.2 $\pm$ 0.1 & 35.2 $\pm$ 0.1 \\
	$19.0-21.0$ & 0.6 & 38.6 $\pm$ 0.1 & -- \\
	$21.0-23.0$ & 0.6 & 38.8 $\pm$ 0.1 & -- \\
	$23.0-25.0$ & 0.6 & 38.1 $\pm$ 0.1 & -- \\
	$25.0-27.0$ & 0.6 & 38.3 $\pm$ 0.1 & -- \\
	$27.0-29.0$ & 0.6 & 37.7 $\pm$ 0.1 & -- \\
	$29.0-31.0$ & 0.6 & 36.7 $\pm$ 0.1 & -- \\
	$31.0-33.0$ & 0.6 & 37.0 $\pm$ 0.2 & -- \\
	$33.0-35.0$ & 0.6 & 35.7 $\pm$ 0.1 & -- \\
	$35.0-37.0$ & 0.6 & 35.7 $\pm$ 0.1 & -- \\
	$37.0-39.0$ & 0.6 & 35.1 $\pm$ 0.2 & -- \\
	$39.0-41.0$ & 0.6 & 34.9 $\pm$ 0.1 & -- \\
	$41.0-43.0$ & 0.6 & 34.0 $\pm$ 0.1 & -- \\
	$43.0-45.0$ & 0.6 & 33.9 $\pm$ 0.2 & -- \\
	$45.0-47.0$ & 0.6 & 33.6 $\pm$ 0.1 & -- \\
	\hline \hline
	\end{tabular}} 
\qquad
\subfloat{\begin{tabular}{cccc}
    \hline \hline
	\multicolumn{4}{c}{Source-HPGe Angle: 25$^\circ$} \\
	\multicolumn{4}{c}{Light Yield Relative Sys. Uncertainty: 9.3\,$\%$} \\
	\multicolumn{4}{c}{Charge Yield Relative Sys. Uncertainty: 12.0\,$\%$} \\
	\hline 
	$E_{r}$ Range & HPGe Res. & Light Yield & Charge Yield \\
	{[}keV] & [keV] & [photons/keV] & [electrons/keV] \\  
	\hline
	$16.0-20.0$ & 0.6 & -- & 34.0 $\pm$ 0.2 \\
	$20.0-24.0$ & 0.6 & -- & 34.9 $\pm$ 0.2 \\
	$24.0-28.0$ & 0.6 & -- & 35.7 $\pm$ 0.1 \\
	$28.0-32.0$ & 0.6 & -- & 36.9 $\pm$ 0.1 \\
	$32.0-36.0$ & 0.6 & -- & 37.4 $\pm$ 0.1 \\
	$36.0-40.0$ & 0.6 & 35.6 $\pm$ 0.2 & 39.2 $\pm$ 0.1 \\
	$40.0-44.0$ & 0.6 & 33.9 $\pm$ 0.2 & 39.9 $\pm$ 0.1 \\
	$44.0-48.0$ & 0.6 & 33.0 $\pm$ 0.1 & 40.9 $\pm$ 0.1 \\
	$48.0-52.0$ & 0.6 & 32.3 $\pm$ 0.1 & 42.1 $\pm$ 0.1 \\
	$52.0-56.0$ & 0.6 & 31.1 $\pm$ 0.1 & 42.9 $\pm$ 0.1 \\
	$56.0-60.0$ & 0.6 & 30.4 $\pm$ 0.1 & 43.6 $\pm$ 0.1 \\
	$60.0-64.0$ & 0.6 & 30.0 $\pm$ 0.1 & 44.5 $\pm$ 0.1 \\
	$64.0-68.0$ & 0.6 & 29.0 $\pm$ 0.1 & 45.3 $\pm$ 0.1 \\
	$68.0-72.0$ & 0.6 & 28.8 $\pm$ 0.1 & 45.7 $\pm$ 0.1 \\
	$72.0-76.0$ & 0.6 & 28.3 $\pm$ 0.1 & 46.0 $\pm$ 0.1 \\
	$76.0-80.0$ & 0.6 & 27.6 $\pm$ 0.1 & 47.0 $\pm$ 0.2 \\
	$80.0-84.0$ & 0.6 & 26.9 $\pm$ 0.1 & 47.6 $\pm$ 0.2 \\
	$84.0-88.0$ & 0.6 & 26.3 $\pm$ 0.1 & 48.1 $\pm$ 0.2 \\
	$88.0-92.0$ & 0.6 & 25.6 $\pm$ 0.1 & 49.2 $\pm$ 0.2 \\
	$92.0-96.0$ & 0.6 & 25.2 $\pm$ 0.1 & 49.7 $\pm$ 0.2 \\
	$96.0-100.0$ & 0.6 & 24.5 $\pm$ 0.1 & 49.2 $\pm$ 0.2 \\
	$100.0-104.0$ & 0.6 & 24.2 $\pm$ 0.1 & 49.7 $\pm$ 0.3 \\
	$104.0-108.0$ & 0.6 & 23.3 $\pm$ 0.2 & 51.2 $\pm$ 0.3 \\
	$108.0-112.0$ & 0.5 & 22.7 $\pm$ 0.2 & 51.0 $\pm$ 0.3 \\
	\hline \hline
	\end{tabular}} 
\end{table*}

\begin{table*}
\caption{Weighted fit of measured light and charge yields in neriX detector during runs I and II for data taken at a mean drift field of 2.32 $\pm$ 0.28\,kV/cm, assuming $g1$, $\eta$, and $G$ values stated in text.}
\label{5500V_table}
\subfloat{\begin{tabular}{cccc}
	\hline \hline
	\multicolumn{4}{c}{Source-HPGe Angle: 0$^\circ$} \\
	\multicolumn{4}{c}{Light Yield Relative Sys. Uncertainty: 7.3\,$\%$} \\
	\multicolumn{4}{c}{Charge Yield Relative Sys. Uncertainty: 5.5\,$\%$} \\
	\hline 
	$E_{r}$ Range & HPGe Res. & Light Yield & Charge Yield \\
	{[}keV] & [keV] & [photons/keV] & [electrons/keV] \\  
	\hline
	$1.0-3.0$ & 0.6 & 28.6 $\pm$ 0.7 & 56.9 $\pm$ 0.3 \\
	$3.0-5.0$ & 0.6 & 30.7 $\pm$ 0.4 & 43.0 $\pm$ 0.2 \\
	$5.0-7.0$ & 0.6 & 34.6 $\pm$ 0.2 & 38.6 $\pm$ 0.1 \\
	$7.0-9.0$ & 0.6 & 35.7 $\pm$ 0.2 & 37.8 $\pm$ 0.1 \\
	$9.0-11.0$ & 0.6 & 36.2 $\pm$ 0.1 & 37.5 $\pm$ 0.1 \\
	$11.0-13.0$ & 0.6 & 35.8 $\pm$ 0.1 & 38.1 $\pm$ 0.1 \\
	$13.0-15.0$ & 0.6 & 35.7 $\pm$ 0.1 & 38.3 $\pm$ 0.1 \\
	$15.0-17.0$ & 0.6 & 35.9 $\pm$ 0.1 & 38.7 $\pm$ 0.1 \\
	$17.0-19.0$ & 0.6 & 35.2 $\pm$ 0.1 & 39.4 $\pm$ 0.1 \\
	$19.0-21.0$ & 0.6 & 34.5 $\pm$ 0.1 & -- \\
	$21.0-23.0$ & 0.6 & 33.7 $\pm$ 0.1 & -- \\
	$23.0-25.0$ & 0.6 & 33.2 $\pm$ 0.1 & -- \\
	$25.0-27.0$ & 0.6 & 32.6 $\pm$ 0.1 & -- \\
	$27.0-29.0$ & 0.6 & 31.5 $\pm$ 0.1 & -- \\
	$29.0-31.0$ & 0.6 & 31.2 $\pm$ 0.1 & -- \\
	$31.0-33.0$ & 0.6 & 30.7 $\pm$ 0.1 & -- \\
	$33.0-35.0$ & 0.6 & 29.7 $\pm$ 0.1 & -- \\
	$35.0-37.0$ & 0.6 & 29.2 $\pm$ 0.1 & -- \\
	$37.0-39.0$ & 0.6 & 28.9 $\pm$ 0.1 & -- \\
	$39.0-41.0$ & 0.6 & 28.5 $\pm$ 0.1 & -- \\
	$41.0-43.0$ & 0.6 & 28.1 $\pm$ 0.1 & -- \\
	$43.0-45.0$ & 0.6 & 27.6 $\pm$ 0.1 & -- \\
	$45.0-47.0$ & 0.6 & 27.4 $\pm$ 0.1 & -- \\
	\hline \hline
	\end{tabular}} 
\qquad
\subfloat{\begin{tabular}{cccc}
	\hline \hline
	\multicolumn{4}{c}{Source-HPGe Angle: 25$^\circ$} \\
	\multicolumn{4}{c}{Light Yield Relative Sys. Uncertainty: 9.3\,$\%$} \\
	\multicolumn{4}{c}{Charge Yield Relative Sys. Uncertainty: 9.1\,$\%$} \\
	\hline 
	$E_{r}$ Range & HPGe Res. & Light Yield & Charge Yield \\
	{[}keV] & [keV] & [photons/keV] & [electrons/keV] \\  
	\hline
	$16.0-20.0$ & 0.6 & -- & 38.6 $\pm$ 0.2 \\
	$20.0-24.0$ & 0.6 & -- & 39.9 $\pm$ 0.2 \\
	$24.0-28.0$ & 0.6 & -- & 41.7 $\pm$ 0.2 \\
	$28.0-32.0$ & 0.6 & -- & 42.5 $\pm$ 0.2 \\
	$32.0-36.0$ & 0.6 & -- & 44.2 $\pm$ 0.2 \\
	$36.0-40.0$ & 0.6 & -- & 45.4 $\pm$ 0.1 \\
	$40.0-44.0$ & 0.6 & 27.9 $\pm$ 0.1 & 46.5 $\pm$ 0.1 \\
	$44.0-48.0$ & 0.6 & 26.9 $\pm$ 0.1 & 47.7 $\pm$ 0.1 \\
	$48.0-52.0$ & 0.6 & 25.8 $\pm$ 0.1 & 48.3 $\pm$ 0.1 \\
	$52.0-56.0$ & 0.6 & 25.0 $\pm$ 0.1 & 49.3 $\pm$ 0.1 \\
	$56.0-60.0$ & 0.6 & 24.5 $\pm$ 0.1 & 50.5 $\pm$ 0.1 \\
	$60.0-64.0$ & 0.6 & 24.2 $\pm$ 0.1 & 50.4 $\pm$ 0.1 \\
	$64.0-68.0$ & 0.6 & 23.2 $\pm$ 0.1 & 51.0 $\pm$ 0.1 \\
	$68.0-72.0$ & 0.6 & 22.4 $\pm$ 0.1 & 52.0 $\pm$ 0.1 \\
	$72.0-76.0$ & 0.6 & 22.3 $\pm$ 0.1 & 52.0 $\pm$ 0.1 \\
	$76.0-80.0$ & 0.6 & 21.8 $\pm$ 0.1 & 52.3 $\pm$ 0.1 \\
	$80.0-84.0$ & 0.6 & 21.4 $\pm$ 0.1 & 53.2 $\pm$ 0.2 \\
	$84.0-88.0$ & 0.6 & 21.2 $\pm$ 0.1 & 53.8 $\pm$ 0.1 \\
	$88.0-92.0$ & 0.6 & 20.6 $\pm$ 0.1 & 54.3 $\pm$ 0.2 \\
	$92.0-96.0$ & 0.6 & 20.3 $\pm$ 0.1 & 53.4 $\pm$ 0.2 \\
	$96.0-100.0$ & 0.6 & 20.0 $\pm$ 0.1 & 54.3 $\pm$ 0.2 \\
	$100.0-104.0$ & 0.6 & 19.7 $\pm$ 0.1 & 55.0 $\pm$ 0.2 \\
	$104.0-108.0$ & 0.6 & 19.4 $\pm$ 0.1 & 54.6 $\pm$ 0.2 \\
	$108.0-112.0$ & 0.5 & 18.6 $\pm$ 0.1 & 55.5 $\pm$ 0.3 \\
	\hline \hline
	\end{tabular}} 
\end{table*}

\begin{acknowledgements}
We gratefully acknowledge support from the National Science Foundation for the XENON100 Dark Matter experiment at Columbia University (Grant No. PHYS09-04220). We also thank Junji Naganoma for his support during data taking, and Patrick de Perio, Qing Lin, and Fei Gao for their useful discussions regarding the analysis.
\end{acknowledgements}


\clearpage
\vspace{-0.12in}
\begin{thebibliography}{99}
\bibitem{Aprile_xenon1t} E.~Aprile \textit{et~al}.~(XENON Collaboration), JCAP \textbf{2016}, 027 (2016).
\bibitem{lz} D.~C. Malling \textit{et~al}.~(LZ Collaboration), arXiv:1110.0103.
\bibitem{Aprile:axion} E.~Aprile \textit{et~al}.~(XENON100 Collaboration), Phys. Rev. \textbf{D 90}, 062009 (2014).
\bibitem{Aprile:dcpaper} E.~Aprile \textit{et~al}.~(XENON Collaboration), Science, Vol. 349, 851 (2015).
\bibitem{Aprile:acpaper} E.~Aprile \textit{et~al}.~(XENON Collaboration), Phys. Rev. Lett. \textbf{115}, 091302 (2015).
\bibitem{Baudis_2013} L.~Baudis \textit{et~al}., Phys. Rev. \textbf{D 87}, 115015 (2013).
\bibitem{Aprile_2012a} E.~Aprile \textit{et~al}., Phys. Rev. \textbf{D 86}, 112004 (2012).
\bibitem{Szydagis_2011} M.~Szydagis \textit{et~al}., JINST \textbf{6}, P10002 (2011).
\bibitem{Manalaysay_2010} A.~Manalaysay \textit{et~al}., Rev. Sci. Instrum. \textbf{81}, 073303 (2010).
\bibitem{Akerib_2016} D.~S.~Akerib \textit{et~al}.~(LUX Collaboration), Phys. Rev. \textbf{D 93}, 072009 (2016).
\bibitem{Lin_2015} Q. Lin \textit{et~al}., Phys. Rev. \textbf{D 92}, 032005 (2015).
\bibitem{Valentine_1994} J.~D. Valentine and B.~D. Rooney, Nucl. Instrum. Methods \textbf{A 353}, 37 (1994).
\bibitem{Rooney_1996} B.~D. Rooney and J.~D. Valentine, Nuclear Science, IEEE Transactions on \textbf{43}, 1271 (1996).
\bibitem{Choong_2008a} W.-S. Choong \textit{et~al}., Nuclear Science, IEEE Transactions on \textbf{55}, 1753 (2008).
\bibitem{Choong_2008b} W.-S. Choong \textit{et~al}., Nuclear Science, IEEE Transactions on \textbf{55}, 1073 (2008).
\bibitem{Plante_2011} G. Plante \textit{et~al}., Phys. Rev. \textbf{C 84}, 045805 (2011).
\bibitem{Aprile_2014a} E.~Aprile \textit{et~al}.~(XENON100 Collaboration), J. Phys. G. \textbf{41}, 035201 (2014).
\bibitem{Aprile_2012b} E.~Aprile \textit{et~al}.~(XENON100 Collaboration), Astropart. Phys. \textbf{35}, 573 (2012).
\bibitem{mcmc} D.~F.-Mackey \textit{et~al}., Pub. Astron. Soc. Pac. \textbf{125}, 306 (2013).
\bibitem{nest_v1} M.~Szydagis \textit{et~al}, JINST \textbf{8}, C10003 (2013).
\end{thebibliography}
\end{document}